\documentclass[oldversion]{aa} % for the traditional abstract
\pdfoutput=1
\usepackage{natbib}
\usepackage{graphicx}
\usepackage{txfonts}
\usepackage{calc}
\usepackage{color}
\usepackage{rotating}
\usepackage[OT2, T1]{fontenc}
\usepackage[USenglish]{babel} % amerikanische Silbentrennung
\begin{document}

\newcommand{\sunrise}{\textsc{Sunrise}}
\newcommand{\carcsec}{$\mbox{.\hspace{-0.5ex}}^{\prime\prime}$}

\title{Comparison of solar photospheric bright points between \sunrise{} observations and MHD simulations}

\author{T.~L. Riethm\"uller\inst{1}
   \and S.~K. Solanki\inst{1,2}
   \and S.~V. Berdyugina\inst{3}
   \and M. Sch\"ussler\inst{1}
   \and V. Mart\'{\i}nez~Pillet\inst{4}
   \and A. Feller\inst{1}
   \and A. Gandorfer\inst{1}
   \and J. Hirzberger\inst{1}
   }

\institute{Max-Planck-Institut f\"ur Sonnensystemforschung (MPS),
           Justus-von-Liebig-Weg 3, 37077 G\"ottingen, Germany
     \and
           School of Space Research, Kyung Hee University,
           Yongin, Gyeonggi, 446-701, Republic of Korea
     \and
           Kiepenheuer-Institut f\"ur Sonnenphysik (KIS),
           Sch\"oneckstr. 6, 79104 Freiburg, Germany
     \and
%           Instituto de Astrof\'{\i}sica de Canarias,
%           C/Via L\'actea s/n, 38200 La Laguna, Tenerife, Spain\\
           National Solar Observatory (NSO), Sunspot, NM 88349, USA\\
           \email{riethmueller@mps.mpg.de}
          }

\date{Received 27 March 2014 / Accepted 30 May 2014}

%__________________________________________________________________
\abstract
{
   Bright points (BPs) in the solar photosphere are thought to be the radiative signatures (small-scale brightness enhancements)
   of magnetic elements described by slender flux tubes or sheets located in the darker intergranular
   lanes in the solar photosphere. They contribute to the ultraviolet (UV) flux variations over the
   solar cycle and hence may play a role in influencing the Earth's climate. Here we aim to obtain a
   better insight into their properties by combining high-resolution UV and spectro-polarimetric
   observations of BPs by the \sunrise{} observatory with 3D compressible radiation magneto-hydrodynamical (MHD) simulations.
   To this end, full spectral line syntheses are performed with the MHD data and a careful degradation is applied to
   take into account all relevant instrumental effects of the observations. In a first step it is demonstrated
   that the selected MHD simulations reproduce the measured distributions of intensity at multiple wavelengths,
   line-of-sight velocity, spectral line width, and polarization degree rather well. The simulated
   line width also displays the correct mean, but a scatter that is too small. In the second step, the properties
   of observed BPs are compared with synthetic ones. Again, these are found to match relatively well,
   except that the observations display a tail of large BPs with strong polarization signals (most likely
   network elements) not found in the simulations, possibly due to the small size of the simulation box.
   The higher spatial resolution of the simulations has a significant effect, leading to smaller and more numerous BPs.
   The observation that most BPs are weakly polarized is explained mainly by the spatial degradation,
   the stray light contamination, and the temperature sensitivity of the Fe\,{\sc i} line at 5250.2\,\AA{}.
   Finally, given that the MHD simulations are highly consistent with the observations, we used the simulations
   to explore the properties of BPs further. The Stokes~$V$ asymmetries
   increase with the distance to the center of the mean BP in both observations and simulations,
   consistent with the classical picture of a production of the asymmetry in the canopy. This is the
   first time that this has been found also in the internetwork. More or less vertical kilogauss magnetic
   fields are found for 98\,\% of the synthetic BPs underlining that basically every BP is associated with kilogauss
   fields. At the continuum formation height, the simulated BPs
   are on average 190\,K hotter than the mean quiet Sun, the mean BP field strength is found to be 1750\,G,
   and the mean inclination is $17^{\circ}$, supporting the physical flux-tube paradigm to describe BPs.
   On average, the synthetic BPs harbor downflows increasing with depth. The origin of these downflows is
   not yet understood very well and needs further investigation.
}


\maketitle

%__________________________________________________________________
\section{Introduction}

   Magnetic fields in the network and in active region plage are often concentrated into strong kilogauss
   field elements \citep{Stenflo1973,Solanki2006,Ishikawa2007}. At high spatial resolution these elements
   appear as bright points \citep[BPs;][]{Berger2001} owing to the inflow of radiation from their walls into
   their evacuated interiors \citep{Spruit1976,Deinzer1984}. Many features of magnetic elements are known
   \citep[see][for an overview]{Solanki1993} and various aspects of the underlying model of magnetic flux
   tubes have been tested, but such tests have generally suffered from the fact that the data were not
   able to spatially resolve magnetic elements.

   This situation has changed with the availability of data recorded by the \sunrise{} balloon-borne
   observatory, which have allowed magnetic elements even in the internetwork quiet Sun to be spatially
   resolved \citep{Lagg2010} and have allowed the internal structure of network magnetic features to be
   probed \citep{MartinezGonzalez2012} providing a good opportunity to revisit BPs and to compare
   their observational properties with predictions from state-of-the-art radiation MHD simulations.

   One motivation to study BPs is their contribution to variations in the total solar irradiance (TSI).
   Around the maximum of the solar activity cycle, the reduction of solar irradiance owing to dark
   sunspots and pores is overcompensated by an increased brightness of the BPs \citep{Froehlich2011,Solanki2013}.
   As a result, the TSI, i.e., the irradiance integrated over all wavelengths, is on average higher during
   maximum solar activity than during minimum \citep{Willson1988}. The TSI variations are only weak over the
   solar cycle, but because 60\,\% \citep{Krivova2006} or even more \citep{Harder2009} of the variations
   in TSI are produced at wavelengths shorter than 4000\,\AA{}, the variations in the ultraviolet (UV)
   can be much more relevant. From recent stratospheric observations we know that the BP contrasts are
   particularly high in the UV \citep{Riethmueller2010}, i.e., the radiative properties of BPs possibly
   play an important role in influencing the Earth's climate. A variation of the UV irradiance changes
   the chemistry of the stratosphere, which can propagate into the troposphere and finally influence
   the climate \citep{London1994,Larkin2000,Gray2010,Haigh2010,Ermolli2012}.

   In addition, there are reasons intrinsic to solar physics why BPs are of interest. First, BPs
   are the most easily visible signatures of strong-field magnetic elements and hence have been widely
   observed \citep[e.g.,][]{Muller1984,Berger1995,Berger2001,Utz2009}. Such magnetic elements carry
   much of the magnetic energy, even if they harbor only a small fraction of the magnetic flux.
   Second, the flux tubes guide magneto-hydrodynamic waves, which could contribute to coronal heating
   \citep{Roberts1983,Choudhuri1993} and may be related to the waves found to run along spicules
   \citep{DePontieu2007}. Additionally, flux-tube motions can lead to field-line braiding and the
   build up of energy, which may be released through nanoflares \citep{Parker1988}.

   The first science flight of the \sunrise{} observatory revealed the very high contrasts of
   BPs in the UV \citep{Riethmueller2010}. Here we follow up on this work by carrying out a more in-depth
   analysis of a quiet-Sun region as observed by \sunrise{} and by comparing the data with numerical
   simulations of three-dimensional radiative magnetoconvection. Hence, we further explore the interplay
   between observation and simulation which has been so fruitful in the past. The MHD simulation data were
   degraded with known instrumental effects that were present during the \sunrise{} observations so that
   they can be compared directly with the observational data. We extend existing studies by considering
   many more observational quantities (intensity at multiple wavelengths, line-of-sight (LOS) velocity,
   spectral line width, and polarization degree) when comparing observational data and MHD simulations,
   which allowed us to test the realism of the MHD simulations far more stringently than by simply comparing
   intensities. We carried out these comparisons in two steps, first for all pixels in the images and later
   restricted to just the pixels identified as lying within BPs. After we were satisfied that
   the simulations give a reasonable representation of the observations, we used the original,
   undegraded simulations to learn more about the BPs and their underlying magnetic features.

%__________________________________________________________________
\section{Observations, simulations, and degradation}
\subsection{Observations}
   The data we used in this study were acquired with the balloon-borne 1~m aperture \sunrise{}
   telescope that flew from Kiruna in northern Sweden to Somerset Island in northern Canada
   in June 2009 \citep{Barthol2011,Solanki2010}. At the flight altitude of roughly 35\,km,
   99\,\% of the air mass was below the observatory, so that the disturbing influence of the
   Earth's atmosphere (seeing) was minimized. A fast tip-tilt mirror, which was controlled by a
   correlating wavefront sensor, reduced the residual pointing jitter and the onboard adaptive
   optics system corrected the images for low order wavefront aberrations \citep{Berkefeld2011}.
   Two instruments were operated simultaneously: the Sunrise Filter Imager
   \citep[SuFI;][]{Gandorfer2011} and the Imaging Magnetograph eXperiment
   \citep[IMaX;][]{MartinezPillet2011}.

   During the time series considered herein, recorded from 23:00 to 24:00 UT on 2009 June 10,
   the telescope pointed to a quiet-Sun region close to the center of the solar disk ($\mu=0.99$).
   The SuFI instrument observed the Sun at the wavelengths 2995\,\AA{} (33\,\AA{} FWHM, 325\,ms exposure time,
   mainly atomic spectral lines), 3118\,\AA{} (8.5\,\AA{} FWHM, 300\,ms exposure time, part of an
   OH band), 3877\,\AA{} (5.6\,\AA{} FWHM, 65\,ms exposure time, CN band), and 3973\,\AA{}
   (1.8\,\AA{} FWHM, 750\,ms exposure time, Ca\,{\sc ii}~H line). Dark-current and flat-field
   corrections were applied to the SuFI data. The images were phase-diversity (PD) reconstructed
   using the wavefront errors retrieved from the in-flight PD measurements via a PD prism in front
   of the camera. The reconstructed data are referred to as level~2 data
   \citep[see][]{Hirzberger2010,Hirzberger2011}. Here we concentrated on the 3118\,\AA{} and the
   3877\,\AA{} bands. The spectra in these bands, taken from the NSO spectral atlas of
   \citet{Kurucz1984} are plotted in the upper two panels of Fig.~\ref{FigSynthSpectra} (black lines)
   along with the filter profiles (green lines).

   The IMaX instrument scanned the Fe\,{\sc i} line at 5250.2\,\AA{} (Land\'e factor $g=3$) in its L12-2 mode,
   i.e., only Stokes~$I$ and $V$ were measured, at twelve scan positions with two accumulations.
   The twelve scan positions were set to $\lambda-\lambda_0=-192.5,...,+192.5$\,m\AA{} relative
   to the center of the average quiet-Sun profile of the line, in steps of 35\,m\AA. The effective
   spectral resolution of IMaX was 85\,m\AA{} (full width half maximum (FWHM) value). The data were
   corrected for dark current and flat field and interference fringes were removed with a manually
   designed Fourier filter. The IMaX data were then reconstructed with the help of the phase diversity
   technique. Additionally, the instrumental polarization and the residual cross talk with intensity
   was removed. Stray light was not removed from either data set. All intensity images were divided
   by the mean quiet-Sun value, $I_{\rm{QS}}$, that was defined as the average of the image.

   Finally, the Stokes~$I$ profiles were fitted with a Gaussian function to retrieve the spectral line
   parameters: LOS velocity and line width. The increased reliability of the retrieved parameters for
   more scanned line positions was the main motivation for using data with twelve scan positions instead
   of five as in the so-called V5-6 mode of IMaX, which was employed in our previous study
   \citep{Riethmueller2010}. Since we were studying BPs, thought to be associated with strong-field
   relatively vertical magnetic features, the Stokes~$Q$ and $U$ profiles were considered to be less
   important for the present work. This assumption is supported by the analysis of \citet{Jafarzadeh2014b},
   who found that BPs extend nearly vertically in height. Figure~\ref{FigSynthSpectra} shows the relevant parts
   of the solar spectrum taken from an NSO spectral atlas \citep{Kurucz1984}. The spectrum centered on the
   Fe\,{\sc i} line at 5250.2\,\AA{} is plotted in the bottom panel. A simulated profile of the 5250.2\,\AA{} line
   (see section~\ref{Simul}) in the presence of an upflow of 5\,km~s$^{-1}$ is overplotted as a dotted red line
   in order to demonstrate that even strong upflows (or downflows) can be safely identified with the L12-2 mode
   of IMaX (see the blue arrows marking the wavelengths sampled by IMaX in L12-2 mode), but can be missed
   or misidentified with only five scan positions (see red arrows). The image quality of the \sunrise{}
   data depended on the gondola's varying pointing stability. The L12-2 data were recorded over
   only a relatively short period of time during the \sunrise{} flight, when the pointing stability was not
   particularly good. Therefore, the image quality of the analyzed L12-2 data is not quite as good as the V5-6 data
   analyzed by \citet{Riethmueller2010}, which has a better spatial resolution. The LOS velocities were
   corrected for the wavelength shift over the FOV caused by the IMaX etalon \citep[see][]{MartinezPillet2011}.
   In this work, negative LOS velocities correspond to upflows.

   We selected the nine data sets acquired at 23:05:08, 23:09:20, 23:20:22, 23:26:09, 23:31:56, 23:36:39,
   23:42:26, 23:47:09, and 23:53:28 UT for an in-depth study from the one-hour time series. The selection
   was done so that the time interval between two consecutive sets was on average five minutes so as to
   give the BPs some time to evolve between two analyzed data sets. For each data set we checked that the
   pointing stability of the gondola and hence the image quality was as good as any among the L12-2 data,
   although it was found to be somewhat lower than of the best V5-6 data.

\subsection{Simulations}\label{Simul}
   The three-dimensional non-ideal compressible radiation MHD simulations considered here were
   calculated with the MURaM code which solves a system of equations consisting of the continuity
   equation, the momentum equation, the energy equation, the induction equation, and the equation
   of state \citep{Voegler2005b}. The radiative energy exchange rate of the energy equation is determined
   by a non-gray radiative transfer module under the assumption of local thermal equilibrium (LTE).
   The equation of state takes into account effects of partial ionization because they influence the
   efficiency of the convective energy transport. Periodic boundary conditions were used in the
   horizontal directions. A free in- and outflow of matter was allowed at the bottom boundary
   of the computational box under the constraint of total mass conservation, while the top boundary
   was closed (i.e., zero vertical velocity). A statistically relaxed purely hydrodynamical simulation
   was used as an initial condition. From tests with different magnetic fluxes, we estimated the
   mean unsigned vertical magnetic flux density of our quiet-Sun observations to correspond roughly
   to a simulation with a starting value of 30\,G and hence a unipolar homogeneous vertical magnetic
   field of $B_{\rm z} = 30$\,G was introduced into the hydrodynamical simulation (see
   section~\ref{CompAllPixels} for a more precise estimate of the mean flux). The simulation was
   run for an additional 3 hours of solar time to reach, and stay for a sufficiently long time, a
   statistically stationary state. Thirty equidistant snapshots covering 141\,min of solar time were then
   used for this study. The data cubes cover 6\,Mm in both horizontal directions with a cell
   size of 10.42\,km (0\carcsec{}014). In the vertical direction they extend 1.4\,Mm with a 14\,km cell size.
   On average, unit optical depth for the continuum at 5000\,\AA{} is reached about 500\,km below
   the upper boundary. To evaluate the dependence of our MHD results on the mean magnetic flux, we also
   calculated ten snapshots each taken from simulation runs with an initial mean unsigned vertical flux
   density of 0\,G (purely hydrodynamical run), 50\,G, and 200\,G, while all other parameters were
   kept identical to the 30\,G run.

   The output of the MURaM code consisted of data cubes of the density, velocity (x, y, z component),
   total energy density, magnetic field (x, y, z component), as well as gas pressure and temperature.
   For a direct comparison with the observations, Stokes profiles had to be derived from these
   data cubes. This was done by a forward calculation with the SPINOR inversion code
   \citep{Frutiger2000a,Frutiger2000b,Berdyugina2003} that uses the STOPRO routines \citep{Solanki1987}
   to compute synthetic Stokes spectra for atomic and molecular spectral lines assuming LTE
   and solving the Unno-Rachkovsky radiative transfer equations \citep{Rachkovsky1962}.
   All spectral line syntheses in this paper were carried out for the center of the solar disk ($\mu=1$).

   Molecular lines were synthesized using the MOL routine library which is employed in the SPINOR
   and STOPRO codes and is based on theoretical computations and results presented by
   \citet{Berdyugina2002,Berdyugina2003,Berdyugina2005}. In particular, they analyzed violet CN and OH
   lines and suggested that these lines can serve as very sensitive diagnostics of temperature and
   magnetic field fluctuations in the solar atmosphere and sunspots using both imaging and
   spectropolarimetry. The CN lines were previously successfully employed for imaging the quiet
   photosphere \citep{Chapman1970,Sheeley1971,Zakharov2005,Zakharov2007,Uitenbroek2006,Uitenbroek2007}
   and for measuring weak entangled solar magnetic fields with the Hanle effect \citep{Shapiro2007,Shapiro2011}.
   The violet OH lines have not yet been broadly employed for solar studies because of the impediment by
   the terrestiral atmosphere. Recently, \citet{Prokhorov2014} have analyzed diagnostic potentials of
   several molecular bands including the CN and OH violet bands using MHD simulations and 3D radiative transfer.

\begin{figure*}
   \centering
   \includegraphics[width=\textwidth]{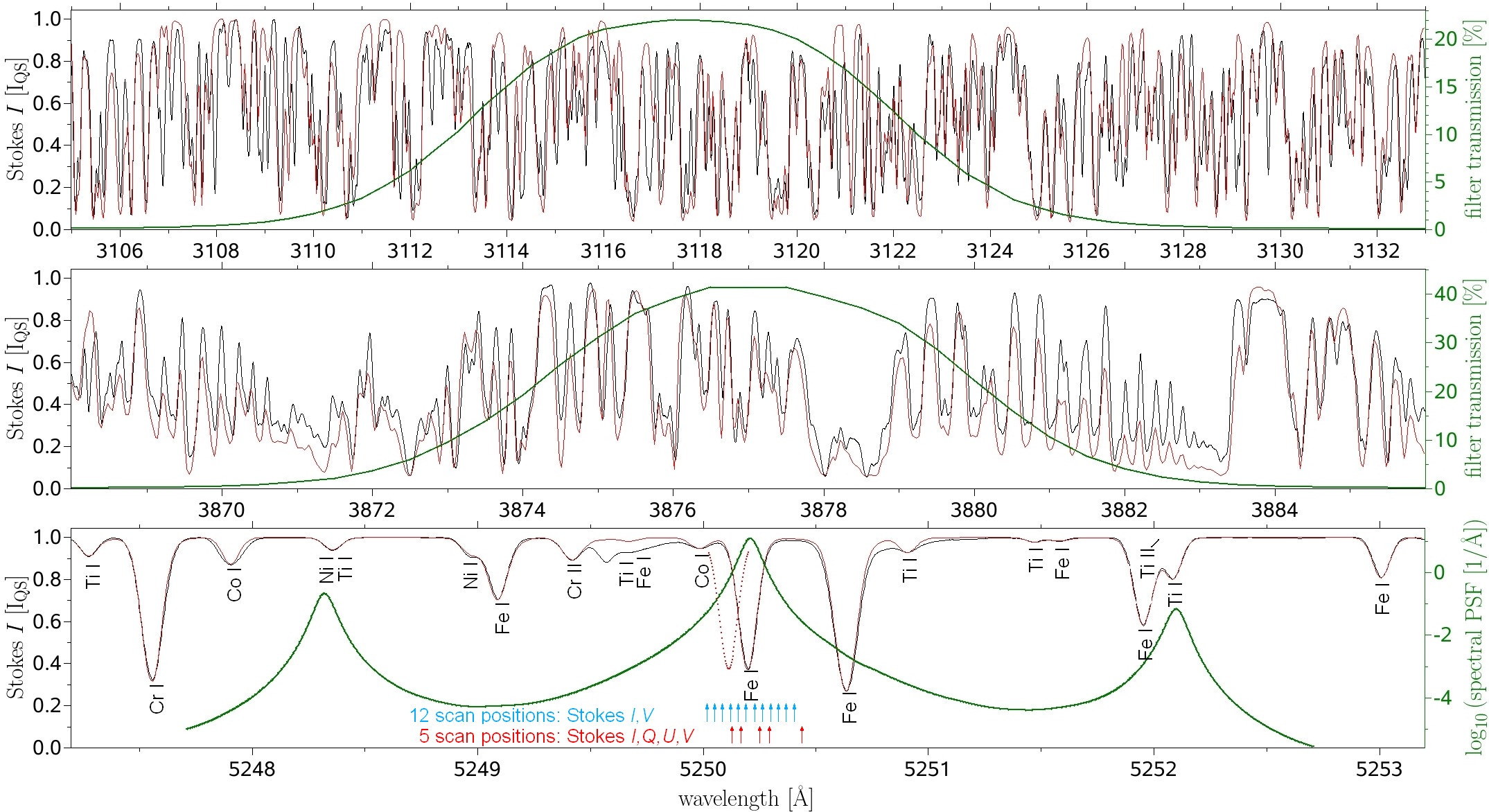}
   \caption{Excerpts from an NSO spectral atlas \citep[black lines;][]{Kurucz1984}, spatially averaged synthetic
   spectra (red lines; see text for details), and instrumental filter profiles (green lines referring to the scale on the right side
   of the figure) for the part of the OH band around 3118\,\AA{} (top panel), the CN band at 3877\,\AA{}
   (middle panel), and the Fe\,{\sc i} line at 5250.2\,\AA{} (bottom panel). The dotted red line in the bottom
   panel simulates the Doppler shifted profile of the Fe\,{\sc i} line at 5250.2\,\AA{} produced by an upflow
   of 5\,km~s$^{-1}$. The red arrows indicate the five scan positions of the IMaX V5-6 mode, while the twelve
   blue arrows mark them for the L12-2 mode.}
   \label{FigSynthSpectra}
   \end{figure*}

   The SuFI spectral range at 3877\,\AA{} includes lines of the CN $B^2\Sigma^+ - X^2\Sigma^+$
   system and the CH $B^2\Sigma^- - X^2\Pi$ system. By comparing synthetic spectra with the
   quiet-Sun NSO atlas spectrum \citep{Kurucz1984}, we have identified 233 CN lines, 36 CH lines,
   and 85 atomic lines. A few lines remained unidentified. In fact, the molecular lines
   completely dominate the chosen spectral range, which includes the CN (0,0) and (1,1) band heads.
   The maximum absorption in these bands is, however, in the wings of the instrument passband
   (see green line in the middle panel of Fig.~\ref{FigSynthSpectra}), while the maximum
   transmission was set at a wavelength where both atomic and molecular lines contribute to
   the absorbtion. We compiled the CN line list using laboratory measured wavelengths and
   by calculating line oscillator strengths from measured molecular constants and band
   oscillator strengths \citep{Knowles1988,Rehfuss1992,Prasad1992,Davis2005,Ram2006}.
   The CH line list was compiled in the same way using wavelengths, molecular constants,
   and band oscillator strengths by \citet{Krupp1974,Bembenek1990}. The CN and CH $B-X$
   systems are strongly perturbed by relatively weak external magnetic fields: the complete
   Paschen-Back effect (PBE) starts at 77\,G and 305\,G for the CN and CH bands, respectively
   \citep{Berdyugina2005}. This implies that even an order of magnitude weaker magnetic field
   will cause noticeable effects, such as line asymmetries and alterations of the line strengths.
   These effects are fully accounted for in our calculations.

   The SuFI 3118\,\AA{} region includes hydroxyl radical OH lines from the $A^2\Sigma^+ - X^2\Pi$ system.
   We have identified 139 OH lines and 399 atomic lines. Thus, the quiet-Sun spectrum in this
   spectral region is dominated by atomic lines. However, at lower temperatures, e.g., in a
   sunspot spectrum, many otherwise weak OH lines become clearly visible. The OH line list was
   compiled using data by \citet{Coxon1980} and \citet{Abrams1994}. The PBE in the OH $A-X$ system
   starts to be noticeable at field strengths stronger than about 200\,G \citep{Berdyugina2003},
   which is taken into account.

   For the spectral range of IMaX we synthesized 20 spectral lines around the Fe\,{\sc i} line at
   5250.2\,\AA{}, all of which could possibly contribute to the
   synthesized Stokes signals owing to the width and secondary peaks of the IMaX spectral point spread function (PSF),
   see green line in the bottom panel of Fig.~\ref{FigSynthSpectra}. The atomic parameters of the 20
   absorption lines around 5250.2\,\AA{} are listed in Table~\ref{AtomicData5250}. Oscillator strengths,
   energies of the lower level, and the term sysmbols have been taken from the Kurucz \citep{Kurucz1995},
   the VALD \citep{Piskunov1995,Kupka2000}, or the NIST \citep{Kramida2012} database, respectively.
   Values in brackets are given only for comparison of the various data sources. Central wavelengths
   and various oscillator strengths are fitted to the NSO spectral atlas.

   Molecular number densities were calculated under the assumption of the chemical equilibrium of
   about 300 molecular species composed of more than 30 atoms, as described in \citet{Berdyugina2003}.
   The solar abundance of iron was taken from \citet{BellotRubio2002}; for all other elements,
   including carbon, nitrogen, and oxygen, we assumed solar abundances according to \citet{Grevesse1998}.

   \begin{table*}
   \begin{minipage}{\linewidth}                         % Minipage wird für Fußnoten innerhalb einer Tabelle gebraucht
   \renewcommand{\footnoterule}{}                       % störende Abgrenzungslinie über den Fußnoten vermeiden
   \caption{Central wavelength $\lambda_0$, oscillator strength $\log{(gf)}$, energy of the lower level $E_l$, and term sysmbol of the lower and upper level of the 20 synthesized spectral lines around the Fe\,{\sc i} line at 5250.2\,\AA{}.}
   \label{AtomicData5250}                               % is used to refer this table in the text
   \begin{tabular}{l l l l l l l l l}                   % 9 left aligned columns
   \hline                                               % inserts single horizontal line
   \noalign{\smallskip}
   Species      & $\lambda_0$ [\AA{}] & $\log{(gf)}$\footnotemark[1]{} & $\log{(gf)}$\footnotemark[2]{}  & $\log{(gf)}$\footnotemark[2]{}  & $\log{(gf)}$\footnotemark[2]{}  & $E_l$ [eV]     & Term$_l$               & Term$_u$               \\
                &                     &              & (Kurucz)      & (VALD)        & (NIST)        &                &                                                 \\
   \hline                                                                                                      
   \noalign{\smallskip}                                                                                        
   Ti\,{\sc i}  & \tt{5247.289}       & \tt{-O.83~}  & \tt{(-O.727)} & \tt{(-O.64~)} & \tt{(-O.727)} & \tt{2.1O3O9~~} & $\mathrm{^{5}F_{3}  }$ & $\mathrm{^{5}F_{2}  }$ \\
   Cr\,{\sc i}  & \tt{5247.566}       & \tt{-1.73~}  & \tt{(-1.64~)} & \tt{(-1.64~)} & \tt{(-1.63~)} & \tt{O.961O39~} & $\mathrm{^{5}D_{O}  }$ & $\mathrm{^{5}P_{1}  }$ \\
   Co\,{\sc i}  & \tt{5247.917}       & \tt{-2.11~}  & \tt{(-2.O7~)} & \tt{(-2.O7~)} & \tt{(-2.O8~)} & \tt{1.7854O4~} & $\mathrm{^{4}P_{1/2}}$ & $\mathrm{^{4}D_{1/2}}$ \\
   Ni\,{\sc i}  & \tt{5248.366}       & \tt{-2.466}  & \tt{(-2.426)} & \tt{(-2.426)} & \tt{(n.a.~~)} & \tt{3.941254~} & $\mathrm{^{3}G_{3}  }$ & $\mathrm{^{3}F_{2}  }$ \\
   Ti\,{\sc i}  & \tt{5248.37O}       & \tt{-1.36~}  & \tt{(n.a.~~)} & \tt{(-1.818)} & \tt{(n.a.~~)} & \tt{1.879~~~~} & $\mathrm{^{3}G_{4}  }$ & $\mathrm{^{3}F_{4}  }$ \\
   Ni\,{\sc i}  & \tt{5248.983}       & \tt{-1.72~}  & \tt{(-2.613)} & \tt{(-2.613)} & \tt{(n.a.~~)} & \tt{3.941254~} & $\mathrm{^{3}G_{3}  }$ & $\mathrm{^{1}F_{3}  }$ \\
   Fe\,{\sc i}  & \tt{5249.1O3}       & \tt{-1.35~}  & \tt{(-1.48~)} & \tt{(n.a.~~)} & \tt{(-1.46~)} & \tt{4.473573~} & $\mathrm{^{3}G_{3}  }$ & $\mathrm{^{3}F_{3}  }$ \\
   Cr\,{\sc ii} & \tt{5249.437}       & \tt{-2.6~~}  & \tt{(-2.426)} & \tt{(n.a.~~)} & \tt{(-2.62~)} & \tt{3.757897~} & $\mathrm{^{4}P_{3/2}}$ & $\mathrm{^{6}D_{5/2}}$ \\
   Ti\,{\sc i}  & \tt{5249.568}       & \tt{-O.863}  & \tt{(-O.793)} & \tt{(n.a.~~)} & \tt{(n.a.~~)} & \tt{3.166432~} & $\mathrm{^{3}P_{1}  }$ & $\mathrm{^{1}D_{2}  }$ \\
   Fe\,{\sc i}  & \tt{5249.682}       & \tt{-2.498}  & \tt{(n.a.~~)} & \tt{(-2.498)} & \tt{(n.a.~~)} & \tt{4.733~~~~} & $\mathrm{^{3}D_{3}  }$ & $\mathrm{^{3}F_{3}  }$ \\
   Co\,{\sc i}  & \tt{5249.997}       & \tt{-O.15~}  & \tt{(~O.32~)} & \tt{(~O.32~)} & \tt{(n.a.~~)} & \tt{4.175373~} & $\mathrm{^{4}G_{5/2}}$ & $\mathrm{^{4}H_{7/2}}$ \\
   Fe\,{\sc i}  & \tt{525O.2O8}       & \tt{-4.938}  & \tt{(-4.938)} & \tt{(n.a.~~)} & \tt{(-4.938)} & \tt{O.121274~} & $\mathrm{^{5}D_{O}  }$ & $\mathrm{^{7}D_{1}  }$ \\
   Fe\,{\sc i}  & \tt{525O.645}       & \tt{-2.21~}  & \tt{(-2.O5~)} & \tt{(n.a.~~)} & \tt{(-2.181)} & \tt{2.198O14~} & $\mathrm{^{5}P_{2}  }$ & $\mathrm{^{5}P_{3}  }$ \\
   Ti\,{\sc i}  & \tt{525O.921}       & \tt{-2.26~}  & \tt{(n.a.~~)} & \tt{(-2.363)} & \tt{(n.a.~~)} & \tt{O.826~~~~} & $\mathrm{^{5}F_{3}  }$ & $\mathrm{^{5}D_{2}  }$ \\
   Ti\,{\sc i}  & \tt{5251.482}       & \tt{-2.81~}  & \tt{(n.a.~~)} & \tt{(-2.541)} & \tt{(n.a.~~)} & \tt{O.818~~~~} & $\mathrm{^{5}F_{2}  }$ & $\mathrm{^{5}D_{1}  }$ \\
   Fe\,{\sc i}  & \tt{5251.595}       & \tt{-2.2~~}  & \tt{(n.a.~~)} & \tt{(-3.O35)} & \tt{(n.a.~~)} & \tt{5.O64~~~~} & $\mathrm{^{5}F_{3}  }$ & $\mathrm{^{6}D_{4}  }$ \\
   Fe\,{\sc i}  & \tt{5251.964}       & \tt{-1.99~}  & \tt{(n.a.~~)} & \tt{(-3.869)} & \tt{(n.a.~~)} & \tt{3.5732187} & $\mathrm{^{1}H_{5}  }$ & $\mathrm{^{5}H_{6}  }$ \\
   Ti\,{\sc ii} & \tt{5252.O19}       & \tt{-2.572}  & \tt{(-2.5O2)} & \tt{(-1.96~)} & \tt{(n.a.~~)} & \tt{2.59O41~~} & $\mathrm{^{2}F_{7/2}}$ & $\mathrm{^{2}F_{5/2}}$ \\
   Ti\,{\sc i}  & \tt{5252.1OO}       & \tt{-2.59~}  & \tt{(-2.448)} & \tt{(-2.448)} & \tt{(-2.448)} & \tt{O.O47969~} & $\mathrm{^{3}F_{4}  }$ & $\mathrm{^{3}F_{3}  }$ \\
   Fe\,{\sc i}  & \tt{5253.O21}       & \tt{-3.8~~}  & \tt{(-3.94~)} & \tt{(n.a.~~)} & \tt{(n.a.~~)} & \tt{2.278758~} & $\mathrm{^{3}P_{2}  }$ & $\mathrm{^{5}P_{1}  }$ \\
   \hline                                               % inserts single line
   \end{tabular}
   \footnotetext{$^1$Adapted value, $^2$Values from the named sources}
   \end{minipage}
   \end{table*}

   For the three considered wavelength ranges, the spatially averaged synthetic spectra
   (calculated for a snapshot with 30\,G average vertical field\footnote{We also compared
   the NSO spectra with averaged synthetic spectra of a purely hydrodynamic snapshot, but found
   only a slight difference in the profiles between 0\,G and 30\,G.}) are shown in Fig.~\ref{FigSynthSpectra}
   (red lines) and compared with the average observed spectra of quiet Sun taken from the NSO atlas
   \citep{Kurucz1984}. The synthetic spectra were convolved with a Gaussian of
   FWHM~=~1.93\,km\,s$^{-1}\sqrt{2/\ln{2}}~\lambda/c$ to fit the spectral resolution of the NSO atlas.
   The spectral PSF of IMaX (green line in the bottom panel) includes the IMaX prefilter and is plotted
   on a logarithmic scale for a better visibility of the secondary peaks at 5248.32\,\AA{} and 5252.10\,\AA{}.

\subsection{Synthetic instrumental effects}
   One of the most sensitive parts of this work was the introduction of synthetic instrumental effects,
   i.e., the degradation of the synthetic data to the same spatial and spectral resolution, to the same
   stray light contamination and noise level as the observed data. The various degradation steps significantly
   influenced the values of the parameters we compared between simulation and observation. We found that only
   if all the relevant effects of the \sunrise{} instrumentation are known to relatively high precision,
   can the comparison between synthetic and observed data be meaningful. All the used degradation steps are
   explained in the following. They were applied in the same order as described below.

\subsubsection{Spectral resolution and sampling}
   The synthetic Stokes profiles output from the SPINOR code have perfect spectral resolution and
   high spectral sampling which had to be reduced to the values of the \sunrise{} instruments. The
   transmission profiles of the SuFI filters as well as the spectral PSF of IMaX (including the pre-filter)
   were measured in the laboratory before the launch of \sunrise{}. The transmission profiles of the two
   SuFI filters (3118\,\AA{} and 3877\,\AA{}) were re-sampled to the wavelength grid points of the
   synthetic intensity profiles. Then the filter transmission profile were multiplied by the intensity
   profile point by point and the products were summed up. This scalar product gave the intensity at a
   spatial pixel of a synthetic SuFI image. In the case of IMaX, the synthetic Stokes profiles were convolved
   with the spectral PSF of IMaX. Finally, Stokes images at the twelve scan positions of the IMaX L12-2 mode
   were retrieved.

\subsubsection{Spatial resolution and residual pointing jitter}\label{SpatResol}
   The spatial resolution of the \sunrise{} data was limited by the 1~m aperture, the considered wavelength,
   and the stability of the gondola's pointing. The theoretical diffraction limit was not fully reached
   during the first science flight of \sunrise{} owing to the residual pointing jitter. From azimuthally
   averaged power spectra of the SuFI and IMaX intensity images, a precise determination of the spatial
   resolution was not possible, but allowed a rough estimate between 0\carcsec{}20 and 0\carcsec{}24
   for the data analyzed here, which are not as highly resolved as data used for earlier publications.
   In the ideal case of a perfect knowledge of the \sunrise{} PSF and no pointing jitter, a spatial
   degradation of the synthesized data would not be needed because we compared with reconstructed
   \sunrise{} data. Theoretically, the deconvolution with the PSF reconstructs the original rms
   contrasts. In the non-ideal case of \sunrise{}, the residual pointing jitter during the observation
   was taken into account by convolving all synthetic Stokes images with a two-dimensional Gaussian
   of FWHM~=~0\carcsec{}23. This FWHM value led to the best match between the rms contrasts of the
   observed and synthesized IMaX continuum images as we found after some tests with different FWHM values.

\subsubsection{Stray light}
   In the case of ground-based solar observations, the atmospheric stray light is a significant part
   of the total amount of stray light. Owing to the flight altitude of approximately 35\,km on average,
   we expect that atmospheric stray light is negligible, so that we have to deal with instrumental stray
   light alone. By observing the solar limb, intensity profiles of the limb could be recorded for the
   considered wavelengths. These solar limb profiles were then compared with the intrinsic limb profiles
   of the Sun taken from the literature \citep{Dunn1968} which allowed the stray light modulation transfer functions (MTFs) to be
   calculated. The stray light MTFs were then multiplied with the synthetic Stokes images in Fourier space.
   Details about the determination of the stray light affecting \sunrise{} data are given by \citet{Feller2014}.

\subsubsection{Noise}
   The noise level of the IMaX Stokes $V$ images was determined at the continuum wavelength (+192.5\,m\AA{}
   offset from the core of the line), which is generally free of $V$ signals. The signals found by
   \citet{Borrero2010} are sufficiently rare not to influence the noise determination significantly.
   The histograms of the nine PD reconstructed Stokes $V$ continuum images we considered here showed a
   clear Gaussian shape with a standard deviation of $3.3 \times 10^{-3} I_{\rm{QS}}$ ($I_{\rm{QS}}$
   is the mean continuum intensity of the quiet Sun).

   The retrieval of the Stokes $I$ noise level was more difficult because here the standard deviation is
   not only determined by the noise, but also by the granulation pattern. For that reason, we determined
   standard deviations for small regions within granules, assuming the signal to be nearly constant
   for such small regions. We found similar values as for Stokes $V$, which confirmed our assumption of small
   intrinsic variations within the small patches considered. Consequently, we added a Gaussian noise with
   $\sigma=3.3 \times 10^{-3} I_{\rm{QS}}$ to all synthetic Stokes images.

\subsubsection{Plate scale}
   The cell size of the simulation data was 10.42\,km and had to be adapted to the pixel size of the IMaX
   observation of 40.10\,km or to that of the SuFI observation of 15.24\,km. Various tests showed
   that the adopted plate scale hardly influenced the parameters we considered and hence did not affect our
   results (mainly because it is significantly smaller than the width of the spatial PSF). For the sake of
   simplicity, we therefore skipped this degradation step in the following study.

\subsubsection{Retrieval of line parameters}\label{ParamRetrieval}
   In the last step, we fitted a Gaussian to the twelve points of the Stokes $I$ profile of the 5250.2\,\AA{}
   line for each pixel to determine the two values, LOS velocity and line width (as FWHM value). The
   Stokes~$V$ profiles were used for calculations of the circular polarization degree, defined as
   \begin{equation}\label{Eq_Mean_CPi}
   \langle p_{\rm{circ}} \rangle = \frac{1}{12} \sum_{i=1}^{12}{\left| \frac{V_i}{I_i} \right|}~,
   \end{equation}
   where the averaging was done over the twelve scan positions of the 5250.2\,\AA{} line. Additionally,
   the area ($\delta A$) and amplitude ($\delta a$) asymmetry of each Stokes~$V$ profile was calculated
   according to \citet{MartinezPillet1997},
   \begin{equation}\label{Eq_StokesVAmplAsym}
   \delta a = \frac{a_b-a_r}{a_b+a_r}~,
   \end{equation}
   \begin{equation}\label{Eq_StokesVAreaAsym}
   \delta A = \rm{sgn(blue)} \frac{\sum_{i=1}^{12}{V_i}}{\sum_{i=1}^{12}{\left| V_i \right|}}~,
   \end{equation}
   where $a_b$ and $a_r$ are the unsigned blue and red lobe amplitudes, and sgn(blue) is $+1$ for positive
   blue lobes and $-1$ for negative ones. The Doppler shifts of the magnetic component are determined from
   the zero-crossing points of the Stokes~$V$ profile. For SuFI data, only the intensity in each filter
   was available.

%________________________________________________________________
\section{Results}

\begin{figure*}
   \centering
   \includegraphics[width=\textwidth]{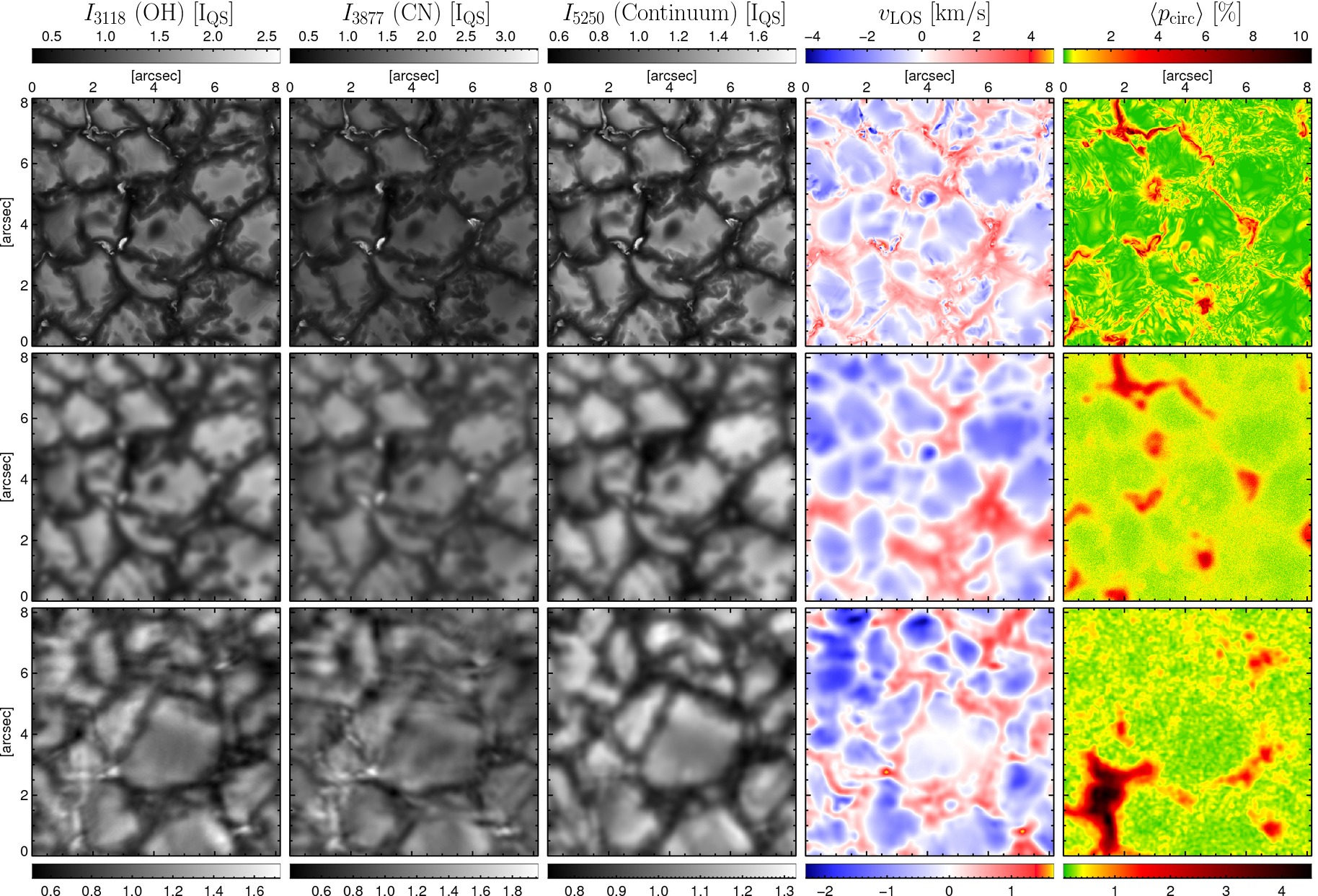}
   \caption{Intensity maps in the wavelength bands 3118\,\AA{} and 3877\,\AA{}, as well as for the continuum at
   5250.4\,\AA{} (first three columns), all normalized to the corresponding mean quiet-Sun intensity
   level, $I_{\rm{QS}}$. The LOS velocity (positive velocities correspond to downflows) and the circular
   polarization degree (see main text for definition) are shown in Cols.~4 and 5, respectively.
   The color bars at the top refer to the undegraded MHD data shown in the first row, while the lower color
   bars apply to the degraded MHD data (second row) and to the data obtained from the \sunrise{}
   observatory (third row).}
   \label{FigImgOverview}
   \end{figure*}

   Figure~\ref{FigImgOverview} contrasts the MHD data with the observational data. The top two
   rows of panels depict a snapshot of the original and the degraded MHD data. The
   bottom panels exhibit an 8.1\arcsec{}$\times$8.1\arcsec{} quiet-Sun region (a subregion of the
   13\arcsec{}$\times$37\arcsec{} common FOV of SuFI and IMaX) as observed with \sunrise{} at
   23:05:08 UT. Several bright granules, separated by darker intergranular lanes, can be seen.
   The undegraded MHD data show many small bright features within the dark lanes. Only the largest
   of these features can be identified as BPs in the degraded MHD data, the smaller ones are smeared
   out by the degradation. The BP contrasts exceed the granulation contrasts in the observed
   OH and CN image, but not for the observed 5250\,\AA{} image, in agreement with the results of
   \citet{Riethmueller2010}. The degraded data look fairly similar to the observations, except that
   some of the observational BP flux concentrations, e.g., the one at position (1.5\arcsec{},2.0\arcsec{}),
   carry more magnetic flux and are bigger than the largest BPs in the degraded simulations. In particular,
   the granulation contrasts in the degraded simulated granulation is very similar to that in the
   observational data at all three wavelengths. This gives us the confidence to proceed with a more
   detailed analysis with the help of histograms of various quantities, plotted in Figs.~\ref{FigOHHist}~to~\ref{FigMean_CPHist}.

\subsection{Simulations versus observations: all pixels}\label{CompAllPixels}
   All pixels of all frames contributed to the histograms plotted in Figs.~\ref{FigOHHist}~to~\ref{FigMean_CPHist}.
   In order to ease comparisons between the similarly shaped histograms, the integral over the
   histograms was always normalized to one.

\subsubsection{Intensity at multiple wavelengths}
\begin{figure}
   \centering
   \includegraphics[width=\linewidth]{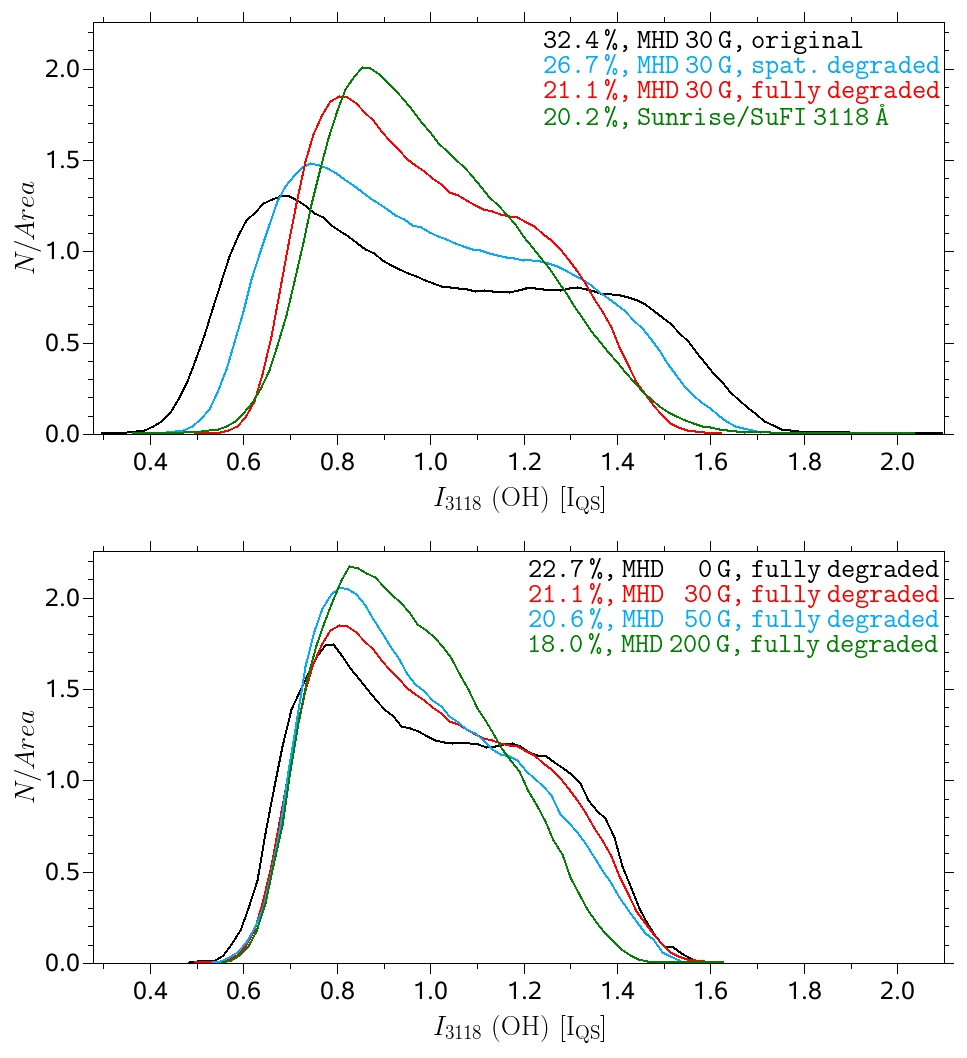}
   \caption{Intensity histograms over all pixels for the OH band data around 3118\,\AA{}. Top panel: The
   black line corresponds to the original 30\,G MHD simulation, the blue line to the spatially degraded data,
   and the red line to the fully degraded simulation data. The green line displays the \sunrise{}
   observations. Bottom panel: Influence of the MHD simulations' mean flux density on the fully
   degraded OH intensity histogram. The black line shows a purely hydrodynamical simulation, i.e., without
   any magnetic field. The mean unsigned vertical flux density was 30\,G for the histogram colored in red,
   50\,G for the blue line, and 200\,G for the green line. RMS contrasts are indicated in the text labels.}
   \label{FigOHHist}
   \end{figure}

\begin{figure}
   \centering
   \includegraphics[width=\linewidth]{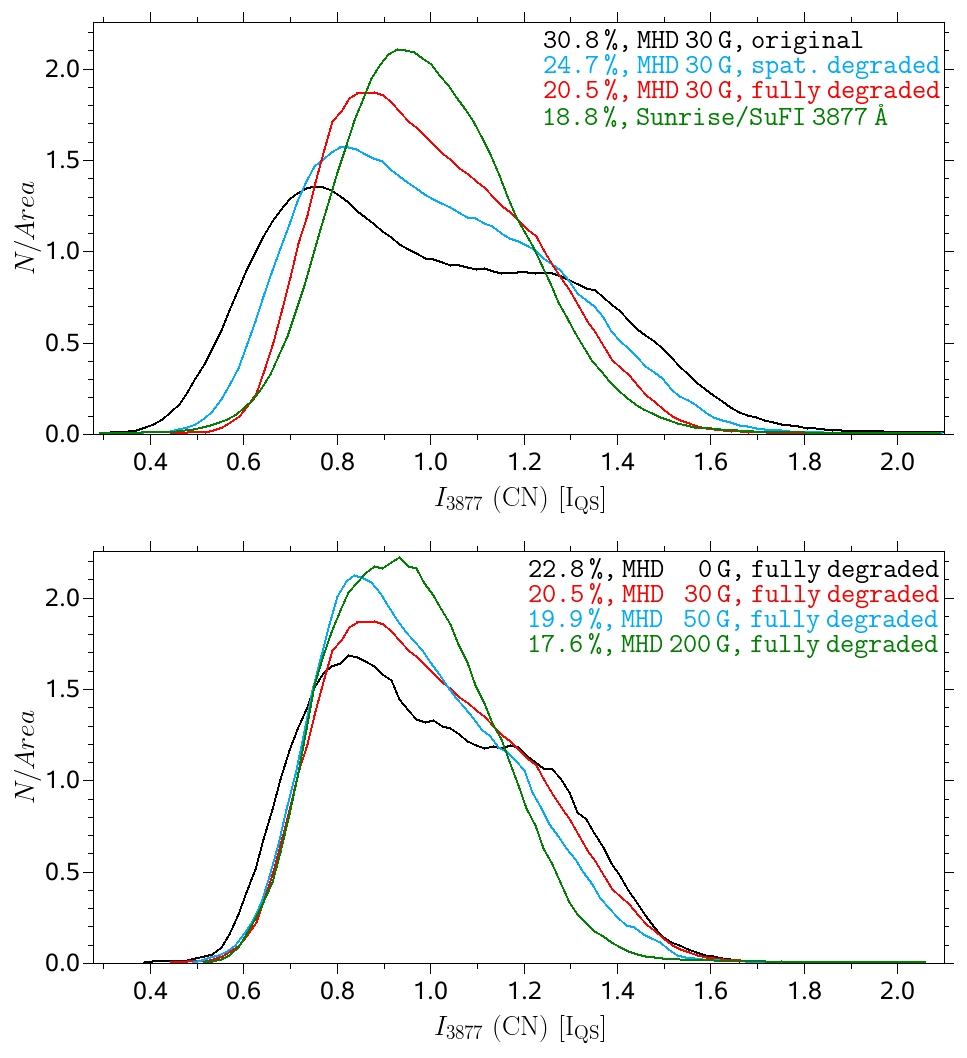}
   \caption{The same as Fig.~\ref{FigOHHist}, but for the CN band data around 3877\,\AA{}.}
   \label{FigCNHist}
   \end{figure}

\begin{figure}
   \centering
   \includegraphics[width=\linewidth]{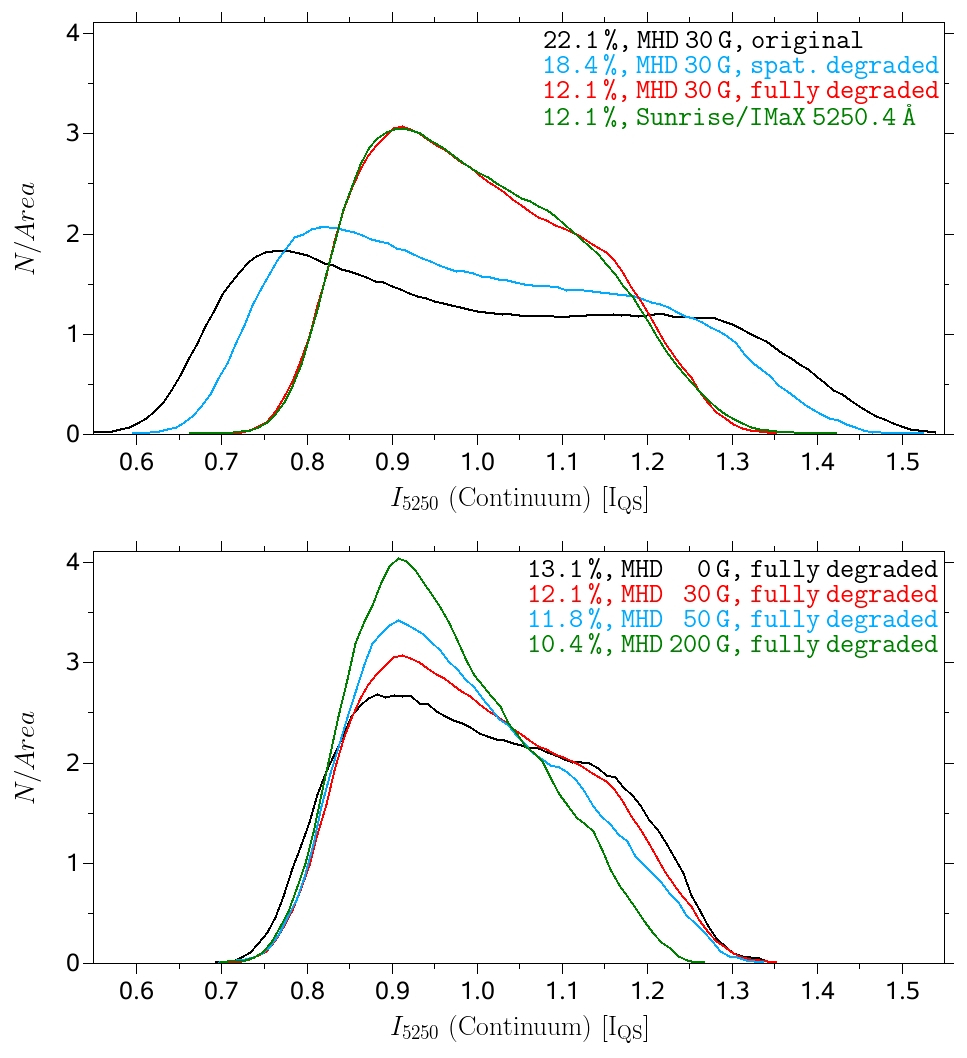}
   \caption{The same as Fig.~\ref{FigOHHist}, but for the continuum at 5250.4\,\AA{}.}
   \label{FigCCTHist}
   \end{figure}

   Figure~\ref{FigOHHist} exhibits histograms of the normalized intensities of the OH data at 3118\,\AA{}.
   The upper panel reveals the influence of the various degradation steps on the histogram of the 30\,G MHD data.
   The histogram of the original MHD data is drawn in black. "Original" means that the computed Stokes~$I$ spectra
   are multiplied with the SuFI filter transmission profile, but no other degradation steps have been applied.
   The blue line represents spectrally and spatially degraded data, while the red line displays the
   fully degraded data, i.e., after spectral, spatial, stray light, and noise degradation. A comparison
   of the black and blue lines shows the influence of the spatial degradation. The noise hardly changes
   the histograms in Figs.~\ref{FigOHHist}~to~\ref{FigFWHMHist} so that the difference between the blue
   and the red line is mainly due to stray light.

   The fully degraded MHD data are then compared with the SuFI observations colored in green. The
   degradation of the simulated data reduces the rms contrast in the OH band from 32.4\,\% down to 21.1\,\%
   which is 0.9\,\% higher than the observational contrast of 20.2\,\%. Although all histograms show a
   certain amount of asymmetry, the histogram of undegraded simulated data is the most extreme and indicates a superposition of two
   populations: the first one consists of intergranular pixels with low intensities, the second one
   contains bright pixels from the granules or from bright points. This superposition is still somewhat
   visible after degradation. The observed histogram does not reveal such a clear superposition of
   two populations.

   The lower panel of Fig.~\ref{FigOHHist} displays OH intensity histograms of fully degraded MHD
   data for different mean unsigned vertical flux densities. The $I_{\rm{QS}}$ for the intensity
   normalization in the considered spectral band is determined from the 30\,G data (closest to our
   observations). The rms contrast is highest in the field-free case, 22.7\,\%, decreasing to 18\,\%
   in the 200\,G simulation (typical mean flux density of a moderate plage region). Without any
   magnetic field, there are no BPs and hence the fraction of the area covered by dark intergranular
   lanes is relatively large. With increasing magnetic flux the number density of BPs increases
   (see discussion of Fig.~\ref{FigBPMean_CPHist}) which reduces the area fraction of the darker
   regions in the intergranular lanes and the rms contrast is reduced as well, but we think that
   this effect plays only a minor role. The decreased contrast reflects mainly the fact that the
   convection is inhibited by a magnetic field \citep{Biermann1941,Spruit1990,Voegler2005a,Kobel2012}.
   We note that the black line in the bottom panel of Fig.~\ref{FigOHHist} not only has more dark pixels
   than the others, but it also has more bright pixels, in particular when comparing with the 200\,G simulation.
   The bright pixels in the field-free case correspond to the edges of granules, which are in some cases
   brighter even than the BPs. The presence of a magnetic field lowers the contrast of the surrounding
   granules. This effect can also be found in the undegraded simulations, where the field-free case
   leads to a contrast of 33.4\,\%, decreasing to 30.8\,\% in the 200\,G simulation.
   The superposition of two populations is most pronounced in the field-free data, but it is not
   visible in the 200\,G data. The shape and width of the observed histogram, as well as its rms contrast
   lies between the degraded histograms of synthetic data with $B_{\rm z} = 50$\,G and $200$\,G.

   The intensity histograms of the CN-band (3877\,\AA{}) are displayed in Fig.~\ref{FigCNHist}. Compared
   with the OH histograms, the superposition of two populations is somewhat less pronounced for the
   simulations. The rms contrast of the simulated 30\,G data is reduced from 30.8\,\% to 20.5\,\%
   after degradation, which is 1.7\,\% higher than the 18.8\,\% contrast of the SuFI data. The main
   difference between the observations and the degraded MHD data is the stronger asymmetry of the synthetic
   histogram. The rms contrast reduces from 22.8\,\% to 17.6\,\% if a 200\,G magnetic field is present.
   The histogram also becomes more symmetric.

   Intensity histograms of the IMaX continuum wavelength $5250.2\,\mathrm{\AA}+192.5\,\mathrm{m\AA} \approx 5250.4\,\mathrm{\AA}$
   are plotted in Fig.~\ref{FigCCTHist}. In Figs.~\ref{FigCCTHist}~to~\ref{FigMean_CPHist}, the term
   "original" MHD data (black lines in the top panels) implies that the Stokes~$I$ spectra are convolved with
   the IMaX spectral PSF, but are not otherwise degraded. The full degradation of the 30\,G MHD data leads
   to a decrease in the rms contrast from 22.1\,\% to 12.1\,\%, which is exactly the observed contrast.
   A very good match between the degraded MHD data and the observations is not only found for the rms
   contrast (which is not a big surprise; see Sect.~\ref{SpatResol}), but also for the shape of the
   histograms which enhances our trust in the applied degradation method. The superposition of two populations
   can be seen for the undegraded data, but it is not so clear for the degraded simulations. The fully degraded
   200\,G MHD data have a 2.7\,\% lower contrast than the corresponding field-free data.

\subsubsection{LOS velocity}
\begin{figure}
   \centering
   \includegraphics[width=\linewidth]{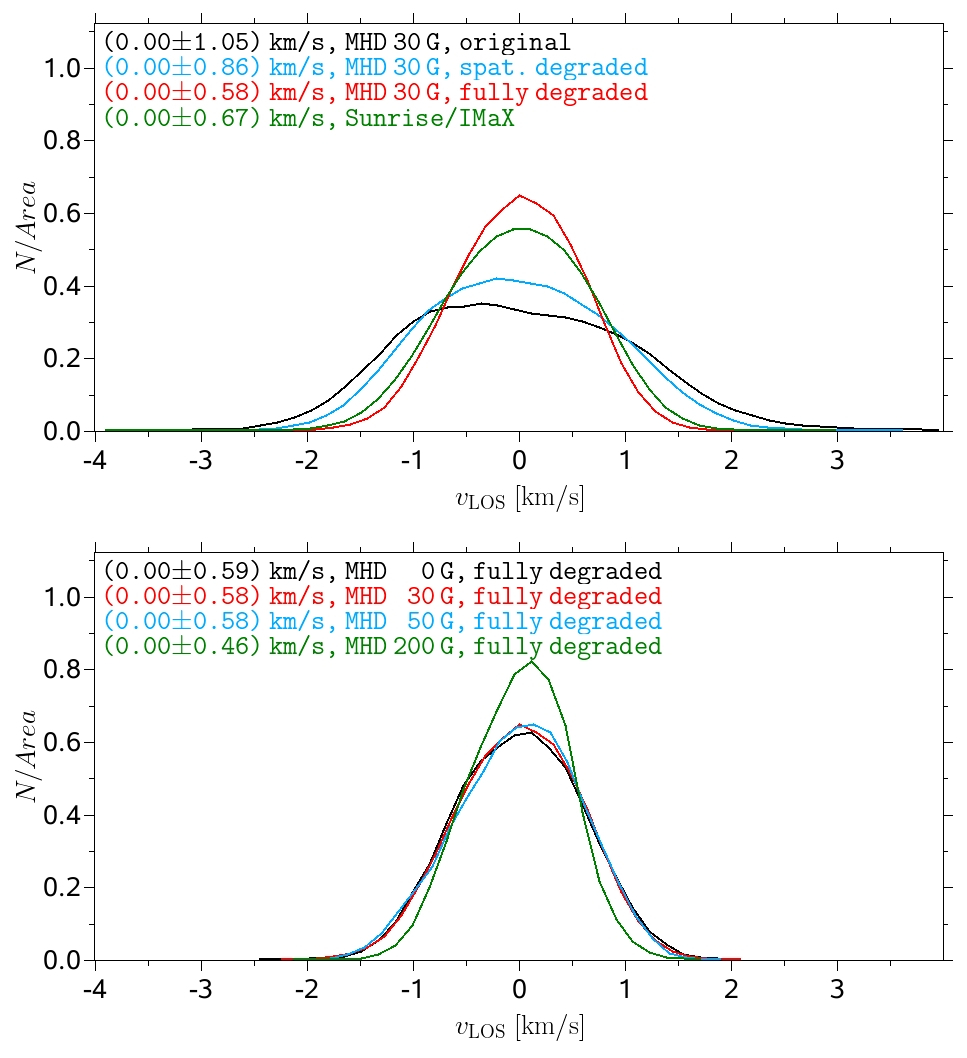}
   \caption{Same as Fig.~\ref{FigOHHist}, but for the LOS velocity as retrieved from a Gaussian fit to
   the Stokes~$I$ profile of Fe\,{\sc i} 5250.2\,\AA{}. Mean values and their standard deviations are
   indicated in the text labels. Negative velocities are upflows.}
   \label{FigLMINHist}
   \end{figure}

   Figure~\ref{FigLMINHist} exhibits histograms of the LOS velocity as determined from the Gaussian fit of
   the twelve scan positions of the Stokes~$I$ profiles of the 5250.2\,\AA{} line. Since an absolute
   wavelength calibration of the \sunrise{}/IMaX data has not been done, we decided to force the
   observed LOS velocities to have the same mean value as the degraded 30\,G MHD data. The standard
   deviation of the 30\,G simulation is reduced by the degradation from 1050\,m~s$^{-1}$ to 580\,m~s$^{-1}$,
   which is close to the standard deviation of the observations of 670\,m~s$^{-1}$. The histograms
   are only weakly asymmetric. Small mean flux densities (0-50\,G) led to almost identical velocity
   histograms. Increasing the average vertical field to 200\,G significantly impedes the convection,
   reducing the standard deviation of the velocities to 460\,m~s$^{-1}$.

\subsubsection{Width of the Fe\,{\sc i} 5250.2\,\AA{} line}
\begin{figure}
   \centering
   \includegraphics[width=\linewidth]{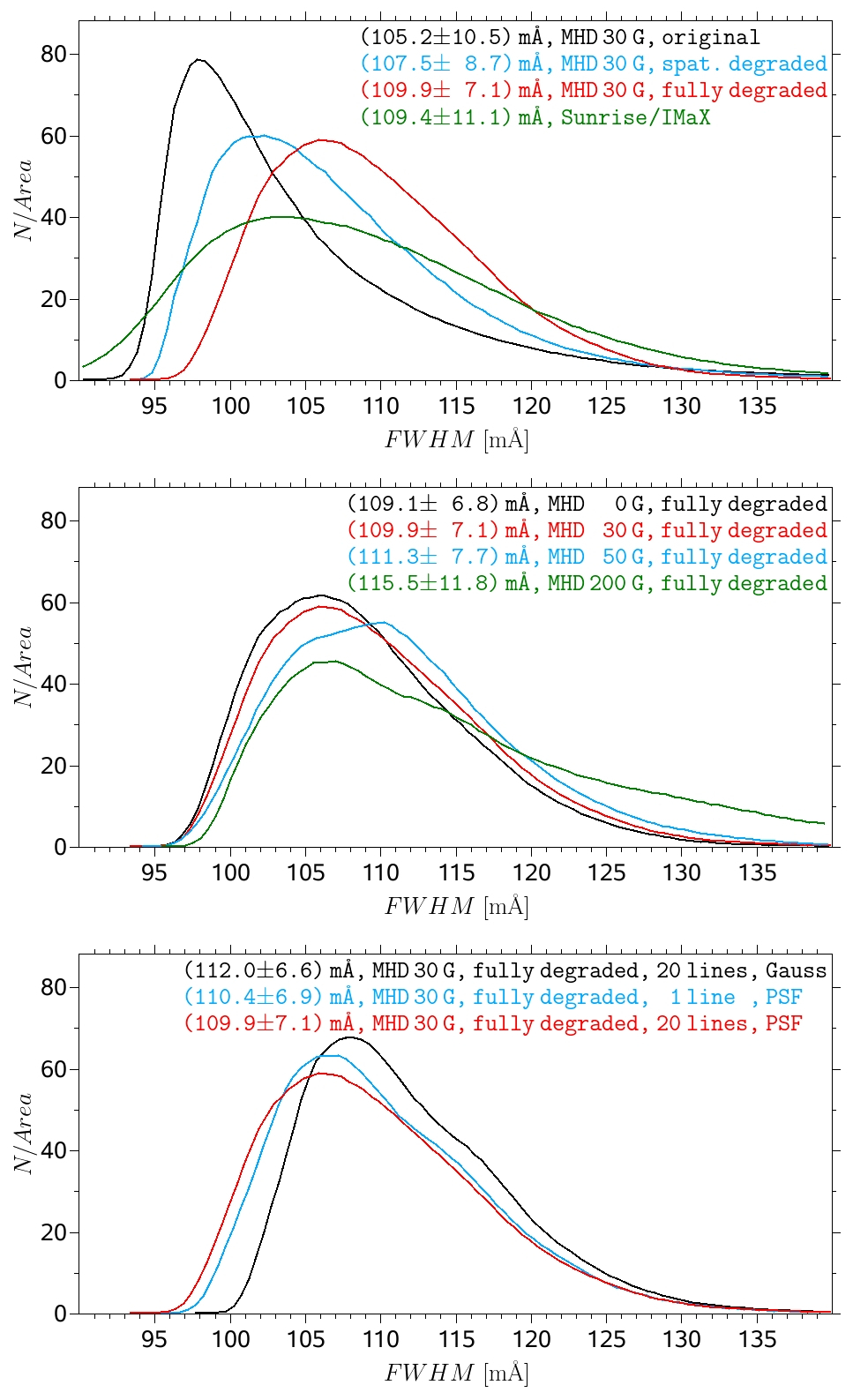}
   \caption{Top and middle panel: Same as Fig.~\ref{FigOHHist}, but for the spectral line width (FWHM)
   as retrieved from a Gaussian fit to the Stokes~$I$ profiles of Fe\,{\sc i} 5250.2\,\AA{}. Bottom panel:
   Influence of different approximations made during the spectral synthesis of the 30\,G MHD data
   on the histogram of the spectral line width. The spectra were fully degraded in all three
   plotted cases. The black line corresponds to the histogram obtained when the filter transmission profile
   was approximated by an 85\,m\AA{} Gaussian instead of the measured spectral PSF. The blue line corresponds
   to a spectral synthesis of only the 5250.2\,\AA{} line, i.e., the 19 neighboring lines were not
   synthesized. For comparison, the red line in the top panel is plotted again and corresponds
   to the 20-line synthesis and spectral degradation with the measured spectral PSF.}
   \label{FigFWHMHist}
   \end{figure}

   Histograms of the 5250.2\,\AA{} line width are displayed in Fig.~\ref{FigFWHMHist}. The degradation
   of the 30\,G simulation data (top panel) causes a shift of the position of the histogram's maximum
   towards larger line widths, coupled with an increase in the width of the histogram, bringing it closer
   to the histogram of the observed values. The histograms of the degraded simulations and the
   observations display a reasonable match for the mean values, but a significant mismatch of their widths,
   i.e., the mean as well as the most common line widths of the simulated profiles are close to the
   observed values, but the scatter of the line widths is clearly larger for the observational data.
   The middle panel of Fig.~\ref{FigFWHMHist} shows how the MHD line width histogram depends on the mean
   flux density. A larger magnetic flux increases the mean value of line widths as well as the standard
   deviation mainly owing to an increased number of larger line width values (partly due to enhanced
   Zeeman splitting). The number of small line width values is hardly influenced by the mean flux
   density and hence none of the additionally considered fluxes matches the observational histogram much
   better than the 30\,G case.

   In the bottom panel of Fig.~\ref{FigFWHMHist} we demonstrate the influence of the broad wings and secondary peaks of the IMaX
   spectral PSF on the degraded 30\,G MHD line width histogram. In particular the approximation of the spectral
   PSF of IMaX by a Gaussian function led to a significantly increased discrepancy between observation and
   simulation. The best match was reached by doing a full 20-line synthesis and using the measured spectral
   PSF for the degradation. We note that convolving with a 85\,m\AA{} Gaussian or restricting the line
   synthesis to just Fe\,{\sc i} 5250.2\,\AA{} left the histograms of all other quantities considered
   in this study practically unchanged.

\subsubsection{Circular polarization degree}
\begin{figure}
   \centering
   \includegraphics[width=\linewidth]{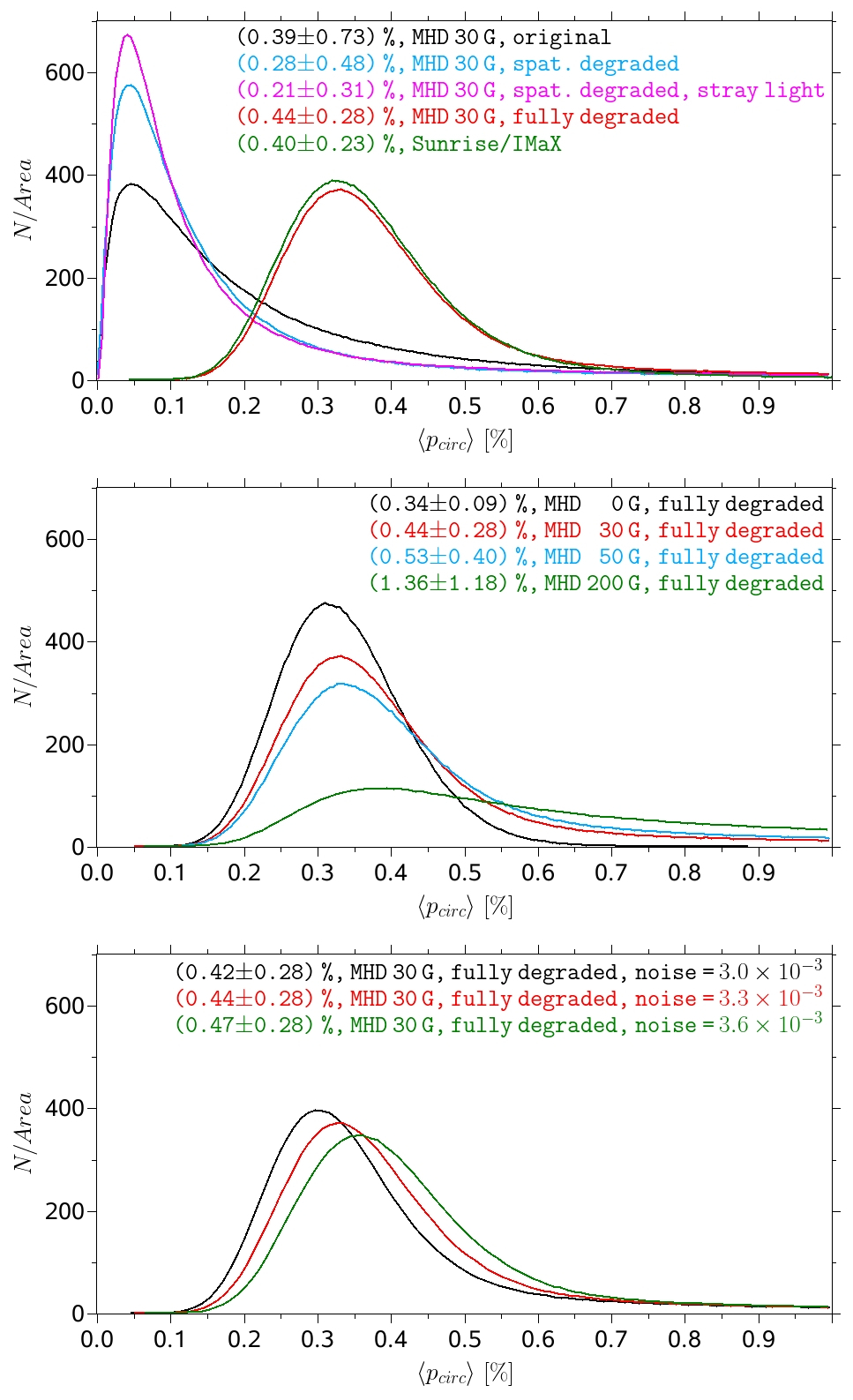}
   \caption{Histograms of the circular polarization degree $\langle p_{\rm{circ}} \rangle$ as retrieved
   from the Stokes~$I$ and $V$ profiles of Fe\,{\sc i} 5250.2\,\AA{} (see Eq.~(\ref{Eq_Mean_CPi})
   for definition). Top panel: The black line corresponds to the original 30\,G MHD simulation, the blue
   line to the spatially degraded data, the magenta line represents the histogram after stray light
   contamination has been added. Finally, the red line corresponds to the fully degraded 30\,G
   simulation. The green line displays the \sunrise{}/IMaX observations. Middle panel: Influence of the
   MHD simulations' mean flux density on the fully degraded polarization histogram. The black line shows
   a purely hydrodynamical simulation, i.e., without any magnetic field. The mean unsigned vertical flux
   density was 30\,G for the histogram colored in red, 50\,G for the blue line, and 200\,G for the green
   line. The noise level was always $3.30 \times 10^{-3}~I_{\rm{QS}}$. Bottom panel: Influence of the
   noise level. The flux density was always 30\,G. The noise levels in units of the mean quiet-Sun
   intensity are printed as text labels.}
   \label{FigMean_CPHist}
   \end{figure}

   Histograms of the circular polarization degree $\langle p_{\rm{circ}} \rangle$ as defined in
   Eq.~(\ref{Eq_Mean_CPi}) are given in Fig.~\ref{FigMean_CPHist}. As in the case of the line widths,
   we found a strong asymmetry in all histograms. This time the match between the degraded 30\,G simulation
   and the observation was remarkably good. The spatial degradation as well as the stray light
   contamination made the histogram narrower and left the maximum position nearly unchanged.
   The influence of the noise was dominant as can be seen by comparing the magenta line (spectral +
   spatial + stray light degradation) and the red line (spectral + spatial + stray light + noise
   degradation). The noise broadens the histogram and shifts the maximum position towards higher
   polarization values. We note that in contrast to $\langle p_{\rm{circ}} \rangle$ the histograms
   of all other parameters considered so far were hardly affected by noise.

   The middle panel of Fig.~\ref{FigMean_CPHist} displays the dependence of the fully degraded MHD
   histogram on the mean vertical flux density. The maximum position and the width of the histograms
   increase with mean flux density. The 0\,G simulation has no intrinsic polarization, so that the
   black line is entirely due to noise that was introduced as part of the degradation. This pure
   noise histogram differs clearly from the observational histogram, so that from such variations
   in the mean MHD flux we estimated that the \sunrise{} data correspond to an average LOS field
   of around 25\,G. (Even the 30\,G simulations shown in the top panel of Fig.~\ref{FigMean_CPHist}
   exhibit a slightly too large mean flux density compared to our observation.) We note, however,
   that all simulations considered here started with an initially homogeneous, vertical and unipolar field.
   A comparison with simulations with a different initial condition (or a different Reynolds number)
   could result in a significantly different estimate of the magnetic field in the observed region
   \citep[see, e.g.,][]{Danilovic2010,Pietarila2010}. The comparison of the histograms of simulated
   SuFI intensities for various mean vertical flux densities with the observations actually suggests
   higher fields of around $50-100$\,G, in better agreement with the above studies. Moreover, the way
   in which the circular polarization degree is normalized -- e.g., division by $I_{\rm{i}}$ as in
   Eq.~(\ref{Eq_Mean_CPi}), division by $I_{\rm{12}}$ (local continuum), or division by
   $\langle I_{\rm{12}} \rangle = I_{\rm{QS}}$, i.e., the mean continuum, see Eq.~(\ref{Eq_Mean_CPc}) -- can
   have a slight influence on the average field strength, estimated from a comparison with simulations.
   Additionally, a possible cross talk from Stokes~$Q$ or $U$ into Stokes~$V$, which cannot be corrected
   because of the lack of Stokes~$Q$ and $U$ signals in the IMaX L12 data, could influence our estimate of
   the mean MHD flux.

   The dependence of the degraded 30\,G MHD polarization histogram on the level of noise, which
   was added to all Stokes images, is shown in the bottom panel of Fig.~\ref{FigMean_CPHist}.
   The $\langle p_{\rm{circ}} \rangle$ histograms for three noise levels between
   $3.00 \times 10^{-3} I_{\rm{QS}}$ and $3.60 \times 10^{-3} I_{\rm{QS}}$ (see text labels) are plotted
   in different colors. Again, the position of the maximum and the width of the histograms increase with
   noise level. Nevertheless, noise level and mean MHD flux density can both be assigned to the
   observations unambiguously since small variations in the mean flux density mainly change the
   amplitude and the width of the histogram, while a small variation in the noise level mainly shifts the
   maximum position. The best-fit noise level was determined to be $\sigma_{\rm{fit}} = 3.30
   \times 10^{-3} I_{\rm{QS}}$, which is somewhat lower than the standard deviation of the observed
   Stokes $V$ continuum signals, $\sigma_{\rm{all}} = 3.77 \times 10^{-3} I_{\rm{QS}}$, suggesting that the
   assumption of a signal-free Stokes $V$ continuum is not entirely correct. Such a signal was found by
   \citet{Borrero2010}, but was restricted to 0.005\,\% of all their spatial pixels. Since we do not
   know how quickly the area covered by these signals increases with decreasing threshold, we use the
   $\sigma_{\rm{fit}}$ value as an approximation of the true noise level.

   In summary, some parameters show better agreement with simulation snapshots with a higher amount
   of initial magnetic flux, others (including the circular polarization amplitude) display the best
   agreement with the 30\,G simulations. Therefore, in the following we will concentrate on comparing
   the observations with this simulation.

\subsection{Simulations versus observations: bright points}\label{CompBPs}
   We now compare BP properties between \sunrise{} observations and the degraded 30\,G simulations.
   From the flux-tube paradigm one would expect kilogauss fields for the BPs and hence strong polarization
   signals, but \citet[][see their Fig.~2, panel h]{Riethmueller2010} find that most of the BPs are
   only weakly polarized. We use the BP detection method applied by \citet{Riethmueller2010} for a direct
   comparison with their results.

   First of all, we need an additional step in the pre-processing of the observational data because
   the IMaX and SuFI instruments differed significantly in their plate scales and there was also a
   slight difference in the plate scales of the various SuFI wavelengths. Therefore, we re-sampled
   all data to the common plate scale of 0\carcsec{}0207 Pixel$^{-1}$ (original plate scale of the
   SuFI 2995\,\AA{} images) via bilinear interpolation. With the help of cross correlation functions
   the two instruments' common field of view (FOV) was identified. Finally, the larger IMaX FOV was
   cropped to the smaller 13\arcsec{}$\times$37\arcsec{} SuFI FOV \citep[see Fig.~1 of][]{Jafarzadeh2013}.

   We then manually identified the BPs' peak intensity in the CN images because of their good visibility
   in that molecular band \citep{Schuessler2003}. The local intensity maximum in an 11$\times$11~pixel patch
   (i.e., 0\carcsec{}22$\times$0\carcsec{}22) surrounding each BP detected at 3877\,\AA{}
   was determined for each of the other wavelengths. This method takes into account that the various
   wavelengths are formed in different atmospheric layers so that inclined features may appear at
   slightly different horizontal positions at different wavelengths \citep[see][]{Jafarzadeh2013}.
   In addition, the dark background (DB) close to the BPs was retrieved. It was defined as the darkest pixel
   whose distance to a BP's local intensity maximum is less than 0\carcsec{}3. We extended the simple
   and manual BP detection method of \citet{Riethmueller2010} by the determination of the BP boundaries
   with the help of a multilevel tracking (MLT) algorithm \citep[see][]{Bovelet2001}. The MLT algorithm
   determined the intensity range of the CN images and subdivided this into 25 equidistant levels.
   Starting with the highest intensity level, all pixels were found whose intensity exceeds this level.
   This led to several contiguous two-dimensional structures, which were tagged with a unique number.
   The obtained structures were extended pixel by pixel as long as the intensity was greater than
   the next lower level. Then the algorithm searched through the whole image again to find all pixels
   whose intensity was greater than the next lower level, which often led to newly detected contiguous
   structures. This procedure was repeated until the minimum intensity level was reached. At the end of the iterations,
   every pixel belonged to exactly one contiguous structure. Finally, contiguous structures which did
   not contain one of the manually detected BPs were ignored and all pixels of the remaining structures
   that had an intensity lower than 50\,\% of the local min-max range were rejected, which led to the
   boundary of the BPs. A more detailed description of MLT, including some illustrative figures,
   has been given by \citet{Riethmueller2008}, who applied the algorithm to the detection of umbral dots.
   The BP boundary detection via MLT was applied to the SuFI intensity images, to the IMaX continuum
   intensity images, and also to the maps of the $\langle p_{\rm{circ}} \rangle$
   (see Sect.~\ref{ParamRetrieval}). We identified 121 BPs in the nine observational data sets.
   This number is limited by the relatively few L12 IMaX data sets available. The corresponding BP number
   density was 0.05 BPs per $\rm{Mm}^2$, which is 1.7 times higher than the value found by
   \citet{Jafarzadeh2013} in the Ca\,{\sc ii}~H channel of \sunrise{}/SuFI. The difference likely
   stems from the restriction to small BPs by \citet{Jafarzadeh2013}.

   We applied the same manual detection method to the degraded CN images of the 30 simulation data sets
   with 30\,G field strength. Here we found 277 BPs (0.26 BPs per $\rm{Mm}^2$). We also detected BPs in
   the 30 undegraded CN images. Owing to the much higher spatial resolution of the undegraded data, we found
   many more BPs there, in total 898 (0.83 BPs per $\rm{Mm}^2$), although many of them were relatively small.
   (Histograms for the BP diameter are discussed in Sect.~\ref{DoubleCellSize}; see also Fig.~\ref{FigBPDiameterHist}.)
   In case of the synthetic data, the boundaries of our manually detected BPs were obtained by applying the MLT algorithm
   to the CN, OH, and 5250.4\,\AA{} intensity images as well as to the maps of the circular polarization degree.
   Furthermore, since we also discuss BP properties which were a direct output of the MHD calculations, such as the
   magnetic field strength, we additionally applied the MLT algorithm to the field strength maps at
   constant optical depth $\log(\tau)=0$ and $\log(\tau)=-2$. The determination of various boundaries
   for the same set of BPs takes into account, e.g., that the magnetic features change in size with height
   and do not always overlap 100\,\% with their brightness enhancements.

\subsubsection{Effective diameter and influence of the MHD cell size}\label{DoubleCellSize}
\begin{figure}
   \centering
   \includegraphics[width=\linewidth]{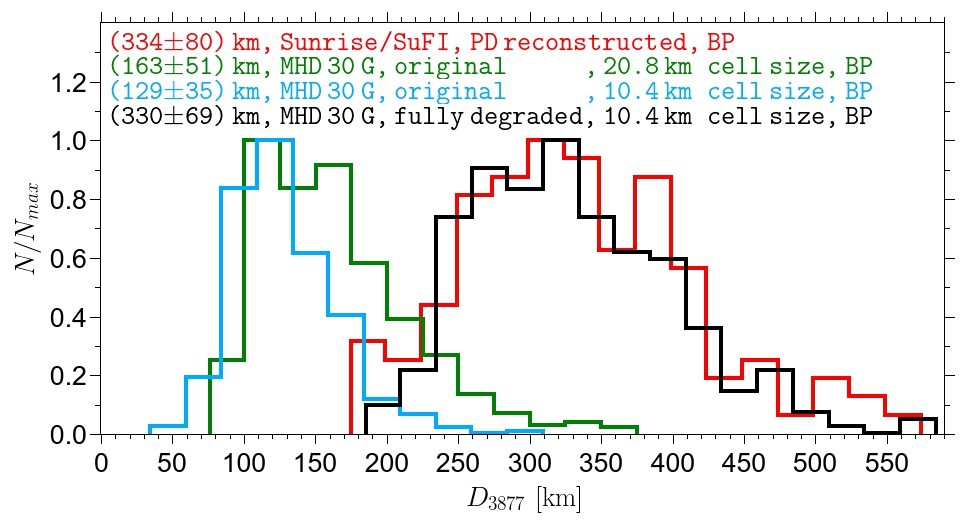}
   \caption{Histograms of the effective BP diameter for the observed BPs (red line), for the undegraded
   MHD BPs simulated with a cell size of 20.8\,km (green line), for the undegraded MHD BPs simulated
   with a cell size of 10.4\,km (blue line), and for the degraded MHD BPs simulated with a cell size
   of 10.4\,km (black line).}
   \label{FigBPDiameterHist}
   \end{figure}

   Figure~\ref{FigBPDiameterHist} compares histograms of the effective BP diameter between the \sunrise{}
   observations and various 30\,G MHD simulations. The diameters were calculated from the BP boundaries
   as determined from the CN intensity images. The observed mean BP diameter of 334\,km (red line) was fairly
   similar to the mean diameter of the degraded synthetic BPs, 330\,km (black line). The influence of the
   degradation can be seen by comparing the black with the blue histogram, representing the undegraded
   MHD simulations. The degradation increased the mean BP diameter from 129\,km to 330\,km.
   If the BP boundaries were determined from intensity images at other wavelengths, we found very similar mean
   diameters for both the observation and the simulation: the observed mean BP diameter is
   ($306 \pm 114$)\,km at 3118\,\AA{} and ($324 \pm 96$)\,km at the 5250.4\,\AA{} continuum wavelength.
   
   The cell size of the MHD grid may also influence some BP properties. We repeated the 30\,G MHD simulation,
   the full spectral line syntheses, and the BP detections for a cell size of 20.8\,km
   while all other parameters were unmodified. We found 475 BPs in the 30 snapshots of
   $6 \times 6\,\mathrm{Mm}^2$ size (number density was 0.44 BPs per $\rm{Mm}^2$) with a mean BP diameter
   of 163\,km (see green line in Fig.~\ref{FigBPDiameterHist}). This dependence of BP diameter and number density
   on the grid size of the simulations is interesting in the sense that it indicates that an even smaller
   grid size could result in smaller BPs. This casts doubts on the claim of \citet{Crockett2010} that they
   resolved essentially all BPs by comparing observed and simulated BP area distributions, since
   the simulations they employed had a grid size of 25\,km, so that true BP sizes are almost certainly
   smaller than was claimed.
   
   Besides the BP diameter and number density we could not find any significant influence of the MHD
   cell size on BP properties: We used the original MHD data to average BP properties over all BPs and
   found for a cell size of 20.8\,km [10.4\,km] a peak intensity at 3118\,\AA{} of $(1.60 \pm 0.39)~I_{\rm{QS}}$
   [$(1.63 \pm 0.45)~I_{\rm{QS}}$], a peak intensity at 3877\,\AA{} of $(2.01 \pm 0.52)~I_{\rm{QS}}$
   [$(2.06 \pm 0.61)~I_{\rm{QS}}$], a peak continuum intensity at 5250\,\AA{} of $(1.27 \pm 0.19)~I_{\rm{QS}}$
   [$(1.29 \pm 0.23)~I_{\rm{QS}}$], a LOS velocity of $(1.28 \pm 0.97)$\,km~s$^{-1}$
   [$(1.25 \pm 1.05)$\,km~s$^{-1}$], a circular polarization degree of $(6.83 \pm 2.03)$\,\%
   [$(6.66 \pm 1.83)$\,\%], and a magnetic field strength at $\log(\tau)=0$ of $(1720 \pm 380)$\,G
   [$(1760 \pm 350)$\,G]. The results in square brackets correspond to MHD simulations with 10.4\,km
   cell size and most of them are discussed in more details in the following text.
   
   We note that we refer to MHD cell sizes of 20.8\,km only in Sect.~\ref{DoubleCellSize}
   (with the green line of Fig.~\ref{FigBPDiameterHist} as the only plotted result). The rest of the
   study stems from 10.4\,km simulations.

\subsubsection{Intensity at multiple wavelengths}
\begin{figure}
   \centering
   \includegraphics[width=\linewidth]{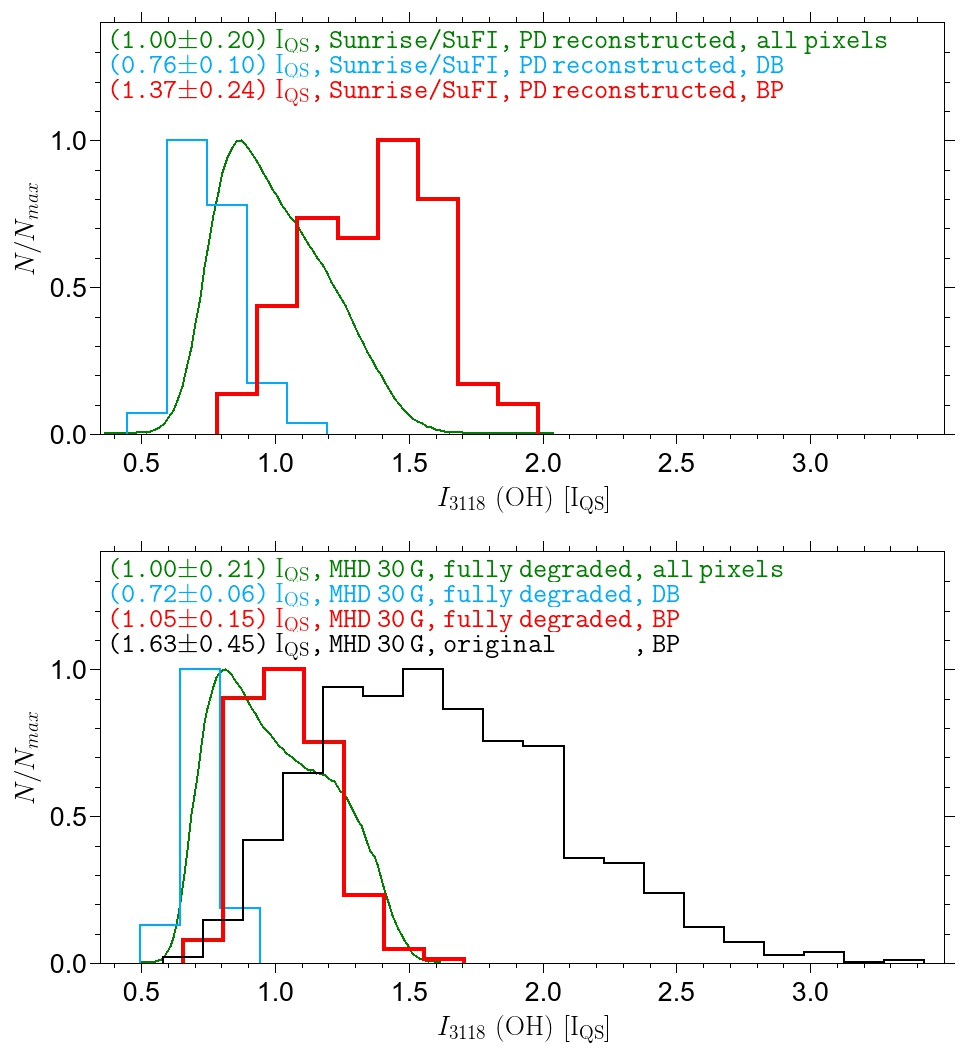}
   \caption{Histograms of the bright point (BP) peak intensity in the 3118\,\AA{} OH band (red lines),
   the intensity histograms of the BPs' dark background (blue lines), and the histograms of all pixels
   (green lines). Mean values and standard deviations are given in the text labels. The top panel
   shows the histograms obtained from the observational data recorded by the SuFI instrument; the bottom
   panel displays the same for the degraded 30\,G MHD simulations. The BP histogram of the undegraded MHD
   data is indicated in the bottom panel by the black line.}
   \label{FigBPOHHist}
   \end{figure}

\begin{figure}
   \centering
   \includegraphics[width=\linewidth]{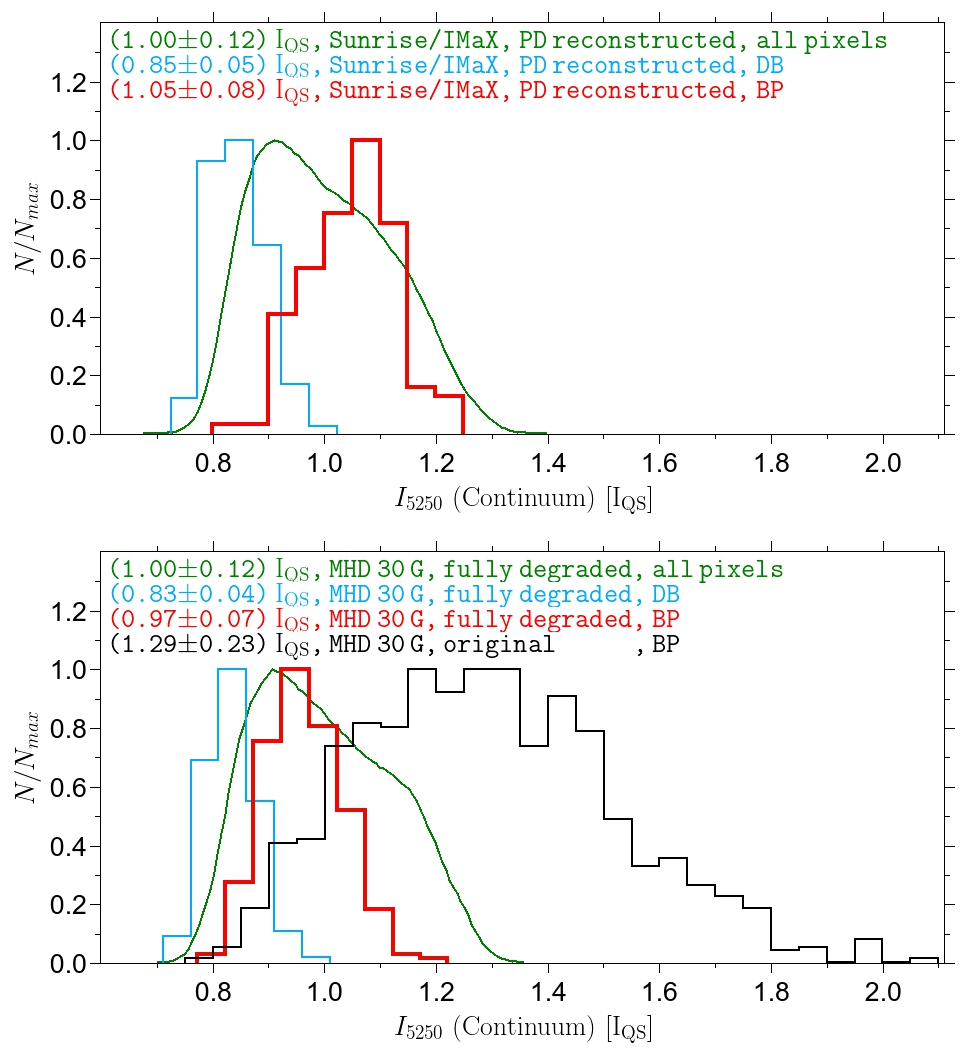}
   \caption{Same as Fig.~\ref{FigBPOHHist}, but for the peak intensity at the 5250.4\,\AA{} continuum.}
   \label{FigBPCCTHist}
   \end{figure}

   Histograms of BP properties are displayed as red lines in
   Figs.~\ref{FigBPOHHist}~to~\ref{FigBPMean_CPHist}. For comparison, the blue lines represent histograms
   of the same parameters in the dark background (DB, defined as the darkest pixel within 0\carcsec{}3
   of each BP's peak position). Likewise for comparison, we plot in green the corresponding
   histograms over all pixels in the images (these green histograms have already been plotted in
   Figs.~\ref{FigOHHist}, \ref{FigCCTHist}, \ref{FigLMINHist}, and \ref{FigMean_CPHist}\footnote{There
   are slight differences between the histograms over all pixels plotted in Figs.~\ref{FigOHHist},
   \ref{FigCCTHist}, \ref{FigLMINHist}, and \ref{FigMean_CPHist} on the one hand, and those in
   Figs.~\ref{FigBPOHHist}, \ref{FigBPCCTHist}, \ref{FigBPLMINHist}, and \ref{FigBPMean_CPHist} on the
   other hand, because the former are obtained from data of the original plate scale and FOV,
   while the latter are produced from the re-sampled data of the common FOV of IMaX and SuFI.}). The histograms of
   Figs.~\ref{FigBPOHHist}~to~\ref{FigBPMean_CPHist} are normalized to their maximum which turned out
   to be more favorable for comparison between histograms of significantly different shapes than
   normalization by their integrals. Figures~\ref{FigBPOHHist}~to~\ref{FigBPMean_CPHist} compare
   the set of 121 BPs detected from the \sunrise{} observations (upper panels) with the set of
   277 BPs detected from the degraded 30\,G MHD simulations (bottom panels). Additionally, the bottom
   panels in Figs.~\ref{FigBPOHHist}~to~\ref{FigBPLMINHist} show the histogram of the 898 BPs that
   we detected in the undegraded MHD data (black lines) in order to illustrate the influence of
   the degradation.

   Histograms of the BP peak intensity (intensity of the brightest pixel of a BP) in the OH band at
   3118\,\AA{} are drawn in Fig.~\ref{FigBPOHHist}. The highest BP intensities coincide
   with the highest intensities found in these images. The observed BPs exhibit a peak intensity
   range of $(0.86-1.97)~I_{\rm{QS}}$ (with a mean value of $1.37~I_{\rm{QS}}$), which is considerably
   higher than the range of $(0.73-1.58)~I_{\rm{QS}}$ with a mean value of $1.05~I_{\rm{QS}}$ obtained
   from the degraded simulations. Hence the observed BPs are brighter. The observed DB histogram is
   also broader than the simulated one.

   Figure~\ref{FigBPCCTHist} shows histograms of the BPs' peak intensity in the continuum at 5250.4\,\AA{}.
   Again, the range covered by the observational BP histogram, $(0.82-1.24)~I_{\rm{QS}}$ is larger than
   that covered by the simulational BP histogram, $(0.80-1.18)~I_{\rm{QS}}$, although the discrepancy is
   not so marked. In addition, we found an almost perfect agreement between observation and simulation
   for the histograms of all pixels and a rather good match for the DB histograms. In contrast to
   the OH band, the highest intensities in the 5250.4\,\AA{} continuum images do not belong to BPs but
   to the brightest parts of granules.

\subsubsection{LOS velocity}\label{BpObsVlos}
\begin{figure}
   \centering
   \includegraphics[width=\linewidth]{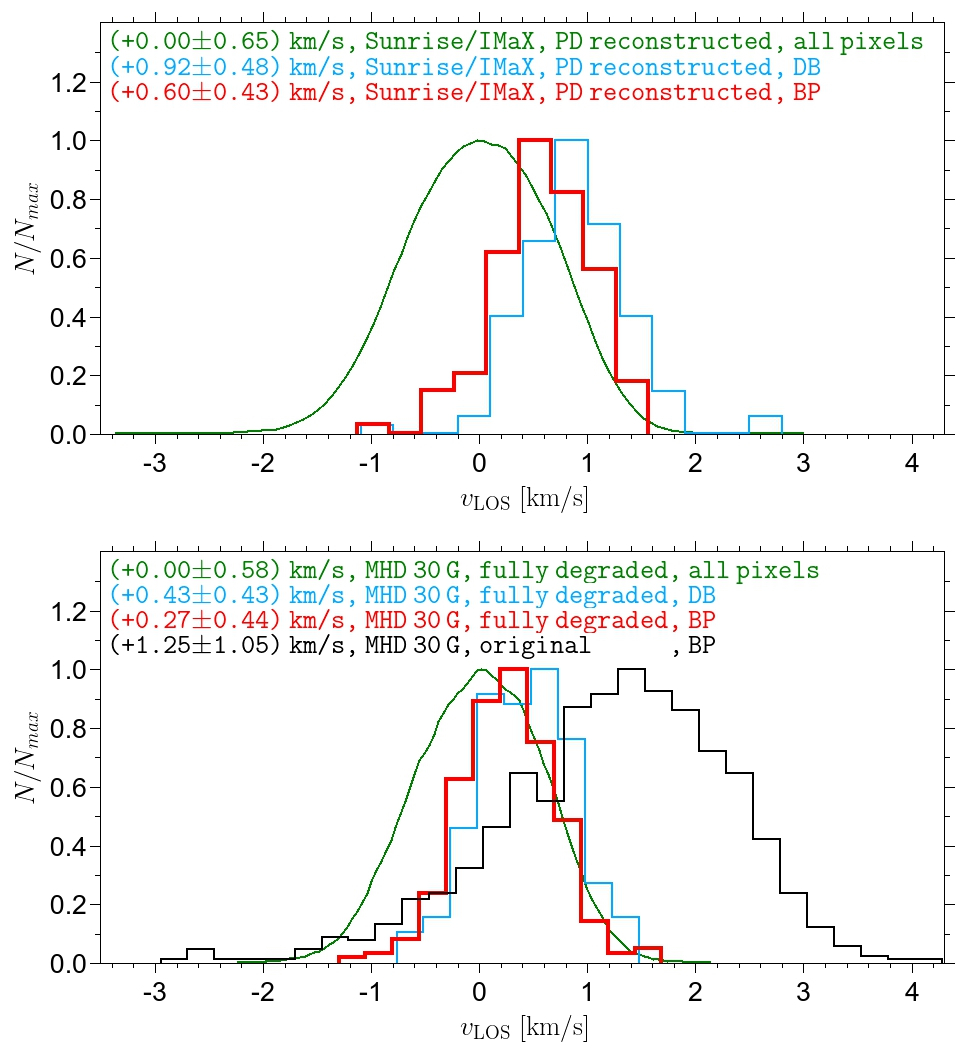}
   \caption{Same as Fig.~\ref{FigBPOHHist}, but for the LOS velocity as retrieved from a Gaussian fit to
   the Stokes~$I$ profile of Fe\,{\sc i} 5250.2\,\AA{}.}
   \label{FigBPLMINHist}
   \end{figure}

   As mentioned above, the LOS velocity was retrieved from a Gaussian fit to the Stokes~$I$ profile of
   Fe\,{\sc i} 5250.2\,\AA{}. Hence the BP boundaries determined from the Stokes~$I$ continuum images were
   used for the spatial averaging of velocities. The average over LOS velocities of all pixels belonging to a
   particular BP was assigned to that BP. The red lines of Fig.~\ref{FigBPLMINHist} exhibit histograms of such
   spatially averaged BP velocities. The standard deviations of the observational histograms agree with
   the degraded simulations, but the observed BPs and their DB display on average larger downflows. Clearly,
   the BPs' LOS velocitiy lies on the downflow side of the distribution for the whole map. The velocities
   of the observed BPs range between $-980$\,m~s$^{-1}$ and 1510\,m~s$^{-1}$, with an average
   downflow of 600\,m~s$^{-1}$. The downflow of the DB is on average 320\,m~s$^{-1}$ stronger. The BPs
   in the simulations show LOS velocities ranging from $-1180$\,m~s$^{-1}$ to 1620\,m~s$^{-1}$ with a
   mean downflow of 270\,m~s$^{-1}$. Here, the downflow of the mean DB is only 160\,m~s$^{-1}$ stronger
   than for the BPs. The degraded simulations lead to a BP velocity distribution which is considerably
   narrower and shows a smaller mean velocity than in the case of the undegraded simulations (black line).
   \citet{Romano2012} also reported that a reduced spatial resolution causes a lower mean BP velocity,
   but a broader BP velocity distribution. \citet{Buehler2014} found that downflows are concentrated at the
   edges of strong-field magnetic features, but not in their cores. At insufficient spatial resolution, however,
   both may get mixed.

\subsubsection{Circular polarization degree}
\begin{figure}
   \centering
   \includegraphics[width=\linewidth]{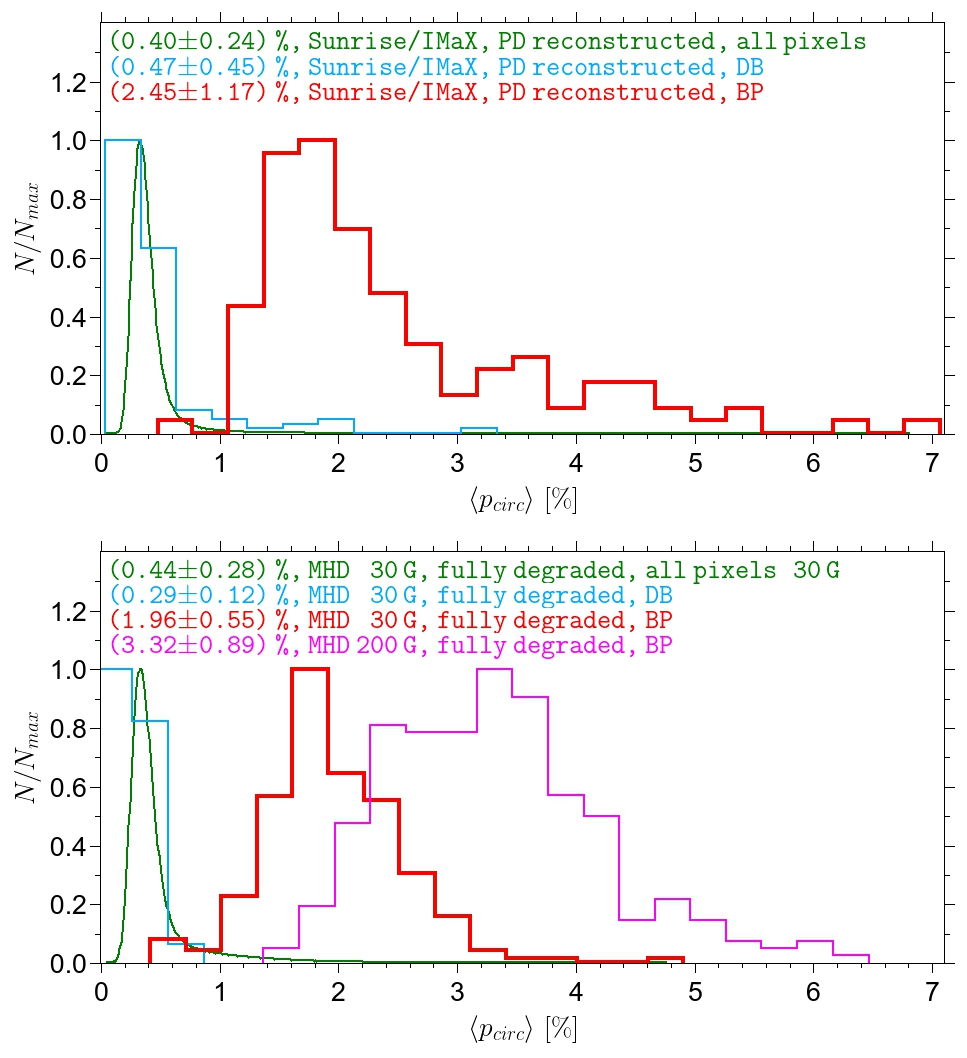}
   \caption{Same as Fig.~\ref{FigBPOHHist} for the spatial peak value of the circular polarization degree.
   Additionally, the magenta line shows the histogram of the BP polarization for the degraded 200\,G
   MHD simulations.}
   \label{FigBPMean_CPHist}
   \end{figure}

   Histograms of the BPs' circular polarization degree (peak polarization within the magnetic structure, i.e.,
   within the BP boundary as determined from the polarization maps) can be seen in Fig.~\ref{FigBPMean_CPHist}.
   A comparison of the upper panel of Fig.~\ref{FigBPMean_CPHist} (IMaX observations with 12 scan positions)
   with Fig.~2, panel~(h) of \citet{Riethmueller2010} (IMaX observations with 5 scan positions) shows that
   most BPs are only weakly polarized, irrespective of spectral sampling.
   The strongest BP polarization degree observed in the V5-6 mode reached 9.1\,\%, while it reached
   only 6.8\,\% in the L12-2 mode. This difference is caused by the fact that we averaged over all twelve
   scan steps in the L12-2 mode and hence over more wavelengths close to the continuum (i.e., low
   polarization signals) than in the V5-6 mode. By interpolating L12-2 Stokes~$V$ to the wavelength positions
   of the V5-6 mode, we obtained similar polarization degrees as found in the V5-6 data by
   \citet{Riethmueller2010}. Furthermore, the noise levels of the two data sets were comparable. We also
   calculated the histogram of the circular polarization degree of the set of V5-6 BPs analyzed by
   \citet{Riethmueller2010} (not shown) and compared it with their histogram of the total polarization
   degree, i.e., $\sqrt{Q^2+U^2+V^2}$ integrated over $\lambda$ in the line, shown in Fig.~2, panel~(h) of
   \citet{Riethmueller2010}. From the similarity of the two histograms we concluded that the Stokes~$Q$
   and $U$ signals are negligible in BPs, so that the lack of $Q$ and $U$ signals in the L12-2 data
   should not affect our conclusions.

   A comparison of the observed and degraded 30\,G simulation BP $\langle p_{\rm{circ}} \rangle$ histograms
   revealed a good agreement for most of the BPs, but a population of BPs showing strong $\langle
   p_{\rm{circ}} \rangle$ is found only in the observations. We can rule out a possible over-reconstruction
   of the IMaX data as the cause of these large $\langle p_{\rm{circ}} \rangle$ because the rms contrasts
   matched rather well. Since we also applied the BP boundary detection via MLT to the circular
   polarization maps, we were able to determine the effective diameter of the polarized features, defined as
   the diameter of a circle of area equal to that within the $\langle p_{\rm{circ}} \rangle$ boundary of
   the BP. We found that 6 of the 121 observed BPs had $\langle p_{\rm{circ}} \rangle > 4.8\,\%$
   (strongest BP polarization of the 30\,G simulation) and their diameter was on average larger by a factor of 1.5
   compared to the mean diameter of the other 115 BPs. Hence, the long tail towards stronger polarization degrees,
   which we found in the observational BP histogram, is caused by large and strongly polarized BPs which
   were not present in the 30\,G simulations. For comparison, we also plotted the histogram of
   $\langle p_{\rm{circ}} \rangle$ of the 285 BPs that we detected in the ten degraded snapshots of our
   200\,G simulations (magenta line in the bottom panel of Fig.~\ref{FigBPMean_CPHist}). There the mean
   polarization degree was 3.32\,\%, with the strongest value being 6.4\,\%, i.e., comparable to the
   strongest observed signals. The number density was 0.79 BPs per $\rm{Mm}^2$.

\subsubsection{Stokes~$V$ asymmetry and zero-crossing velocity}
\begin{figure}
   \centering
   \includegraphics[width=\linewidth]{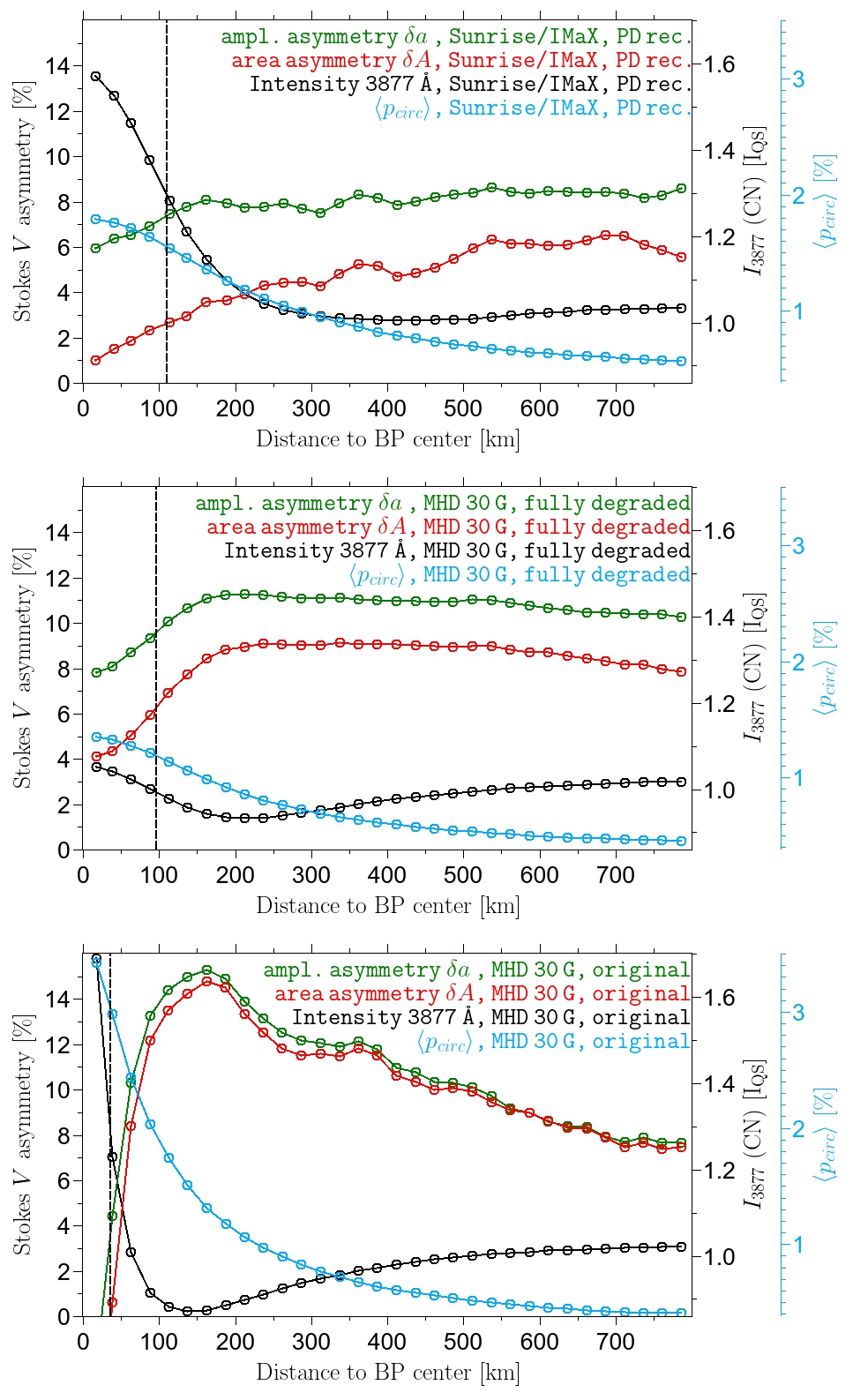}
   \caption{Stokes~$V$ amplitude (green lines) and area (red lines) asymmetries, intensity in the 3877\,\AA{} CN band
   (black lines), and circular polarization degree (blue lines) of the reconstructed IMaX observations (top panel),
   degraded (middle panel), and original (bottom panel) 30~G MHD simulations as a function of the distance
   to the BP center. The dashed black lines display the half width half maximum value of the mean intensity profile.}
   \label{FigStokesVAsym}
   \end{figure}

   In a next step, we azimuthally averaged the Stokes~$V$ asymmetries and zero-crossing velocities of all pixels
   at roughly the same distance to the BP center and plotted these mean values as a function of the distance to
   the center of the BP. Only Stokes~$V$ profiles having a two-lobe shape with a signal-to-noise ratio of at least 3
   for each of the two lobes contributed to the averaging. We restricted the distance range to 800~km since the
   number of pixels fulfilling this condition decreases considerably farther out. Figure~\ref{FigStokesVAsym}
   displays the Stokes~$V$ asymmetries. On average, amplitude and area asymmetry exhibit low values in the
   core of the BPs and increase quickly up to a radial distance of about 160-180~km \citep[see also][]{Shelyag2007}.
   For larger distances, the asymmetries of the original MHD data decrease slowly again, while they more or less
   saturate in the case of the degraded simulations and observations. Even if both kinds of Stokes~$V$ asymmetry
   show slightly larger values for the degraded simulation than for the observation, the qualitative behavior
   is quite similar. Interestingly, area and amplitude asymmetry show nearly identical values in the undegraded
   simulations, i.e., the fact that the observations and the degraded simulations lead to area asymmetries
   that are on average lower than the corresponding amplitude asymmetries is only caused by instrumental effects.
   
   For comparison, we also plotted the mean BP profile for the normalized intensity of the
   CN data at 3877\,\AA{} and for the circular polarization degree in Fig.~\ref{FigStokesVAsym}. (All pixels,
   and not only those having a clear two-lobe Stokes~$V$ profile, contributed to the calculation of the mean BP intensity
   and $\langle p_{\rm{circ}} \rangle$ profiles.) The highest intensities are found in the core of the BP, while
   the lowest values are associated with the interganular lanes. Farther out the intensity increases again, but only
   weakly for the observations. As already retrieved from the histograms in Figs.~\ref{FigBPOHHist} and \ref{FigBPCCTHist},
   the BP intensities of the observations are higher than for the degraded simulations.   
   The FWHM values of the mean BP intensity profile, 220\,km for the observations, 190\,km for the degraded simulations,
   and 70\,km for the original simulations, are a measure of the size of the mean BP brightness structure that is independent
   of any BP boundary detection. These values are smaller than the averages obtained from Fig.~\ref{FigBPDiameterHist}.
   A comparison suggests that the BP boundaries returned by the MLT method correspond roughly to the e-folding width of the intensity.
   
   Additionally, we plotted the mean circular polarization degree and found a monotonic but more gentle decrease with radial distance.
   The FWHM values of the mean polarization structure, 430\,km for the observations, 390\,km for the degraded simulations,
   and 190\,km for the original simulations, are roughly twice as big as the mean brightness structure (compare, e.g., the large
   flux patch at position (1.5\arcsec{},2.0\arcsec{}) in the $I_{\mathrm{3877}}$ and $\langle p_{\rm{circ}} \rangle$ maps of the
   bottom row of Fig.~\ref{FigImgOverview}). That larger flux concentrations are more diffuse than their brightenings was already
   reported by \citet{Berger2001} for their spectropolarimetric observations with the 0.5~m SVST telescope on La Palma and is
   confirmed by our study for the \sunrise{} observations as well as for the MHD simulations. Since the mean BP is much larger in
   polarization than in brightness even in the original MHD data, this is most likely an intrinsic effect and cannot be caused by a pure
   PSF effect similar to the one pointed out by \citet{Title1996}. We expected that the reason for this is related to the expansion of
   flux tubes with height. In locations of the magnetic canopy the Stokes~$V$ profile is formed higher up, where $B \neq 0$, than the
   Stokes~$I$ profile which may get its main contribution from below the canopy. This is particularly true for the broadband 3877\,\AA{}
   SuFI channel. Other possibilities may be that only part of the strong-field magnetic features are bright.

\begin{figure}
   \centering
   \includegraphics[width=\linewidth]{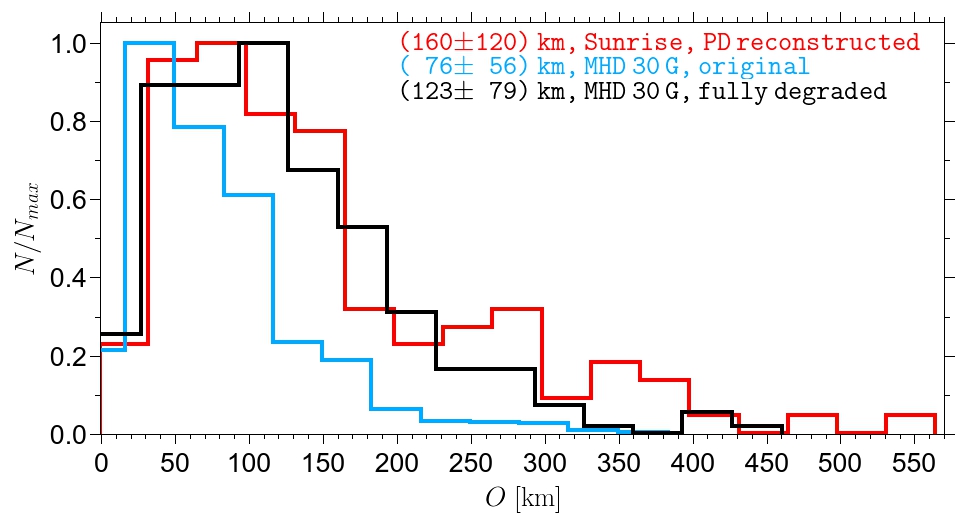}
   \caption{Histograms of the offset between the BPs' peak in CN intensity and the peak in the circular polarization degree,
   $O$, for the observed BPs (red line), the undegraded MHD BPs (blue line), and for the degraded MHD BPs (black line).}
   \label{FigOffs_I388_Mean_CPi}
   \end{figure}

   The maximum brightness of a magnetic feature need not be co-spatial with its maximum magnetic field. This could be
   the case if the brightness is largest near the walls of a flux tube, but its $B$ has its peak in the center of the flux tube.
   By how much the peak in the CN intensity is offset from the peak in the circular polarization degree can be seen
   in Fig.~\ref{FigOffs_I388_Mean_CPi} for the observed and synthesized set of BPs. Mean values and their standard
   deviations are given in the text labels. The histograms show a peak at lower offsets and a moderate decrease towards
   higher offsets. For half of the undegraded synthetic BPs, the offset is smaller than 62\,km, which is less than 50\,\%
   of its effective MLT diameter of 129\,km.

\begin{figure}
   \centering
   \includegraphics[width=\linewidth]{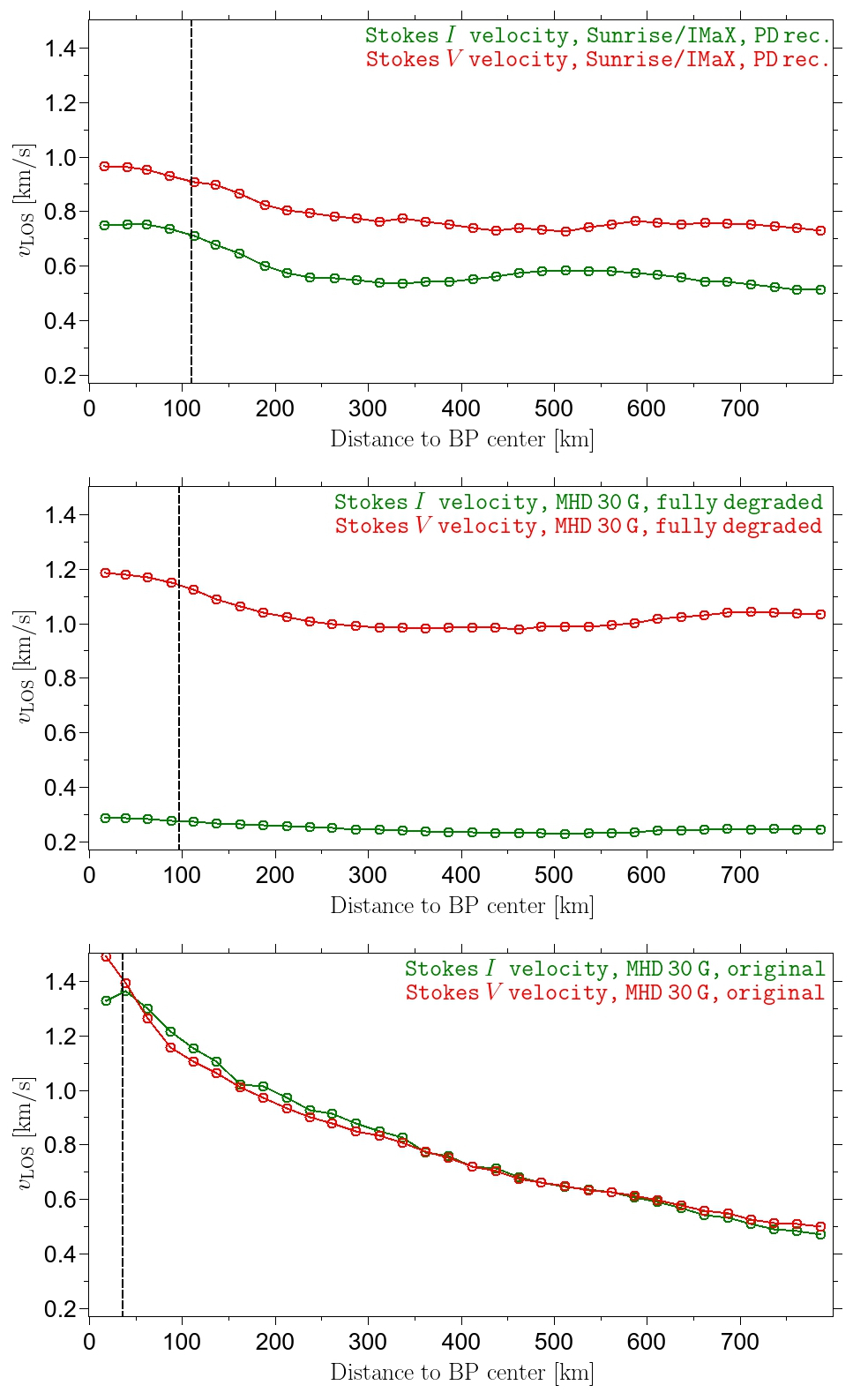}
   \caption{LOS velocity as retrieved from a Gaussian fit to the Stokes~$I$ profile (green lines) and from
   the Stokes~$V$ zero crossing (red lines) of the reconstructed IMaX observations (top panel),
   degraded (middle panel), and original (bottom panel) 30~G MHD simulations as a function of the distance
   to the BP center. The dashed black lines are the same as in Fig.~\ref{FigStokesVAsym}.}
   \label{FigStokesVZeroCross}
   \end{figure}

   Finally, the LOS velocity from the Stokes~$V$ zero-crossing and the Stokes~$I$ Doppler velocities
   are displayed in Fig.~\ref{FigStokesVZeroCross} as a function of the radial distance. Owing to the
   small MHD cell size of 10.4\,km, most of the BPs are well resolved in the original MHD data,
   so that Stokes~$I$ and $V$ lead to rather similar velocity values there. The downflows of about
   1.4\,km~s$^{-1}$ in the BP center decrease continually. Even for the largest shown radial distances,
   we found remarkable downflows of roughly 500\,m~s$^{-1}$ which could possibly be affected by the fact that
   BPs are often located in regions where several intergranular lanes join up, so that azimuthal
   averages at these larger distances can be considerably influenced by intergranular lanes.
   
   The spatial smearing as part of the degradation of the MHD data mixes magnetic and non-magnetic components
   within the resolution element, so that the Stokes~$I$ velocities are significantly decreased by the
   degradation. The Stokes~$V$ zero-crossing velocities do not undergo this degradation because they only represent the
   magnetic component. However, the drop in the MHD Stokes~$I$ velocities due to degradation is so strong
   that the relatively good match between observation and degraded simulation found for the Stokes~$V$ velocities
   cannot be found for the Stokes~$I$ velocities. To a minor degree this can be explained by the fact that
   the observations show more clustering of the magnetic field and the mixture of magnetic and non-magnetic
   components due to spatial smearing acts mainly at the border of larger flux patches, but is not
   so significant for their inner parts \citep[see also][]{Buehler2014}. An analysis in which the degradation
   of the MHD data was applied step by step revealed that the stray light contamination is the main contributor
   to the strong Stokes~$I$ velocity drop. This becomes understandable considering that the fit of the observed
   solar limb profiles \citep[which led to the stray light MTF applied in this study; see][]{Feller2014} could not
   disentangle the stray light from the remaining pointing jitter of the gondola. Hence the determined stray light
   MTF must be seen as an upper limit since it also contains a jitter component. Unfortunately, the jitter varied
   widely over the \sunrise{} mission, so that our method of degradation, for all its shortcomings, is the best
   currently available approach for the \sunrise{} data.
   
\subsection{Why are most bright points weakly polarized?}\label{ExplWeakBPs}
   According to Fig.~\ref{FigBPMean_CPHist} most BPs, both synthetic and observed, are only weakly
   polarized at the resolution of \sunrise{}/IMaX. The same was noted by \citet{Riethmueller2010}.
   This weak polarization could have a number of causes. Either most BPs are associated with
   intrinsically weak fields, contrary to standard flux-tube theory \citep[][see below]{Spruit1976},
   or they are very highly inclined, nearly horizontal, also contrary to expectations for
   strong fields \citep[][a highly evacuated flux tube anchored at one end is driven to be nearly
   vertical by buoyancy]{Schuessler1986}. Alternatively, they could be spatially unresolved at the
   spatial resolution reached by \sunrise{}, or the weak Stokes~$V$ could be caused by thermal
   weakening of Fe\,{\sc i} 5250.2\,\AA{} in BPs. Of course, some combination of these effects
   may also be acting. We searched for the cause by analyzing the simulation data.

   Bright points are often modeled by nearly vertical slender flux tubes. In the flux-tube model, only magnetic
   field strengths in the kilogauss range can explain the brightnesses that are observed in BPs. The field
   increases the magnetic pressure which leads to an evacuation inside the tube and hence a depressed
   optical depth unity surface. The lateral inflow of heat through the walls of the flux tube makes it
   hot and bright.

   To determine what polarization signals can be expected for kilogauss fields, we synthesized Stokes
   profiles for a standard atmosphere, the HSRASP \citep{Chapman1979}, assigned a zero velocity and a
   height independent field strength of 1\,kG. The synthetic profiles were then convolved with the
   spectral PSF of IMaX and the Stokes~$V$ values at the twelve scan positions of the IMaX L12-2 mode
   were used to calculate $\langle p_{\rm{circ}} \rangle$ according to Eq.~(\ref{Eq_Mean_CPi}).
   A value of 10.71\,\% was obtained. This value is significantly higher than the mean value of
   $\langle p_{\rm{circ}} \rangle$ of 1.96\,\% obtained from the fully degraded data.

\begin{figure}
   \centering
   \includegraphics[width=\linewidth]{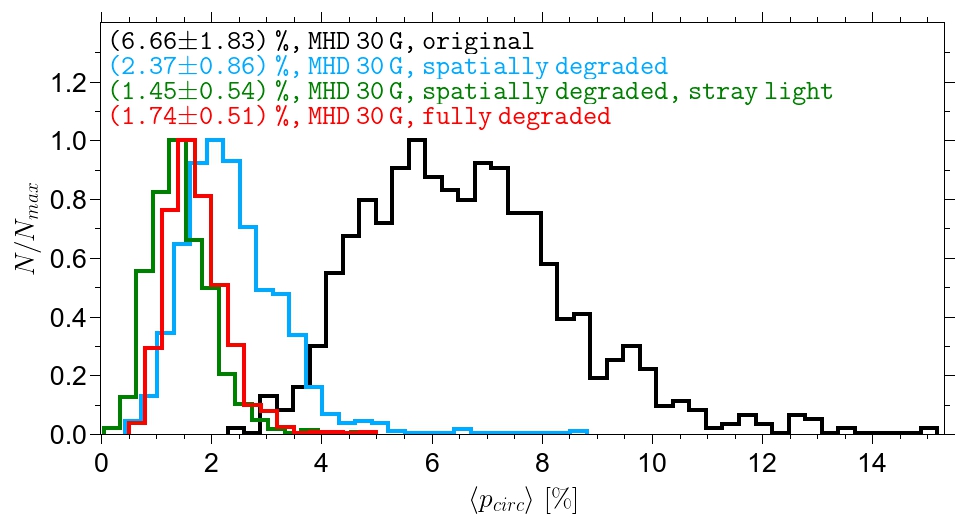}
   \caption{Influence of the various degradation steps on the histogram of the BP circular polarization
   degree. The degradation was applied to the 30\,G MHD simulations. The black line corresponds
   to the data that were only spectrally degraded. An additional spatial degradation results in the blue
   histogram. The green line also includes the effect of stray light contamination, while the fully
   degraded BP signals, i.e., including noise, give the histogram colored in red.}
   \label{FigBPMean_CP2Hist}
   \end{figure}

   To analyze the influence of the various degradation steps on the mean BP polarization, we used
   the set of 898 BPs that we detected in the undegraded CN images for the original\footnote{Again,
   "original" means only spectrally degraded, because this step was part of our synthesis. No
   further degradation steps had been applied at this stage.} MHD data and determined their peak
   $\langle p_{\rm{circ}} \rangle$ values (see black line in Fig.~\ref{FigBPMean_CP2Hist}) for the data
   that were spectrally and spatially degraded (blue line), for the additionally stray light contaminated
   images (green line), and the fully degraded data (red line in Fig.~\ref{FigBPMean_CP2Hist}).
   The spatial degradation reduced the mean BP polarization from 6.66\,\% down to 2.37\,\%. The stray light
   led to a further reduction down to 1.45\,\% and the noise increased the mean value to 1.74\,\%.
   In contrast to the histograms of all pixels (top panel of Fig.~\ref{FigMean_CPHist}), here the noise
   was not the main contributor, but rather the spatial degradation, because the BPs are small (see
   Fig.~\ref{FigStokesVAsym}) and their $\langle p_{\rm{circ}} \rangle$ values are much higher than
   the noise level.

   We note that the red line in the bottom panel of Fig.~\ref{FigBPMean_CPHist} and the red line
   in Fig.~\ref{FigBPMean_CP2Hist} are not identical. Both histograms were retrieved from the
   same fully degraded $\langle p_{\rm{circ}} \rangle$ maps, but the lower panel of
   Fig.~\ref{FigBPMean_CPHist} was calculated for the 277 BPs that were detected in the degraded
   CN images, while Fig.~\ref{FigBPMean_CP2Hist} displays $\langle p_{\rm{circ}} \rangle$ of the
   898 BPs detected from the undegraded CN images. These contained many small BPs and hence led
   to a smaller mean value of 1.74\,\% compared to 1.96\,\% for the 277 BPs detected in the
   degraded CN images.

\begin{figure*}
   \centering
   \includegraphics[width=\textwidth]{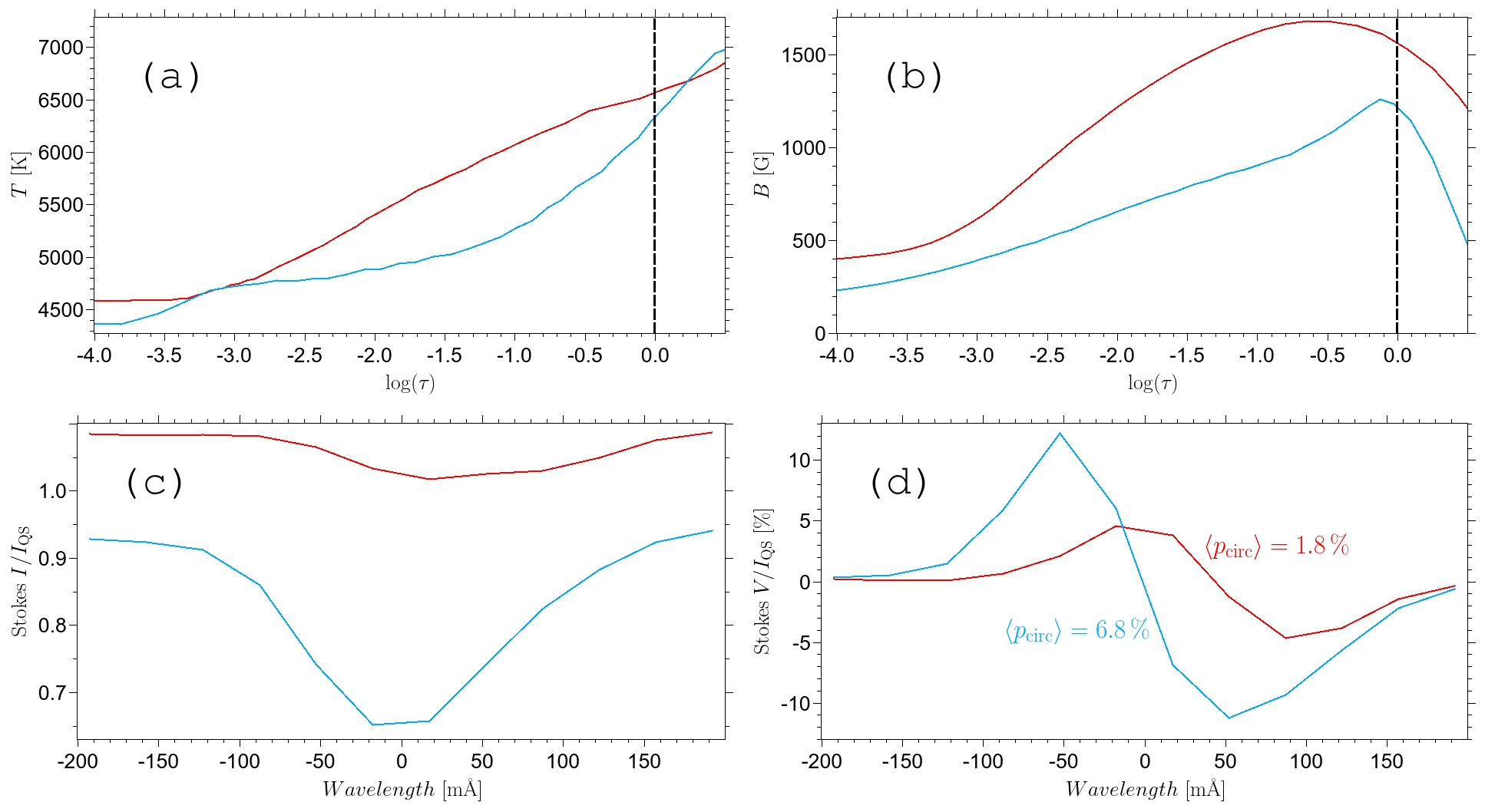}
   \caption{Demonstration of a weak polarization signal owing to the strong temperature sensitivity of the
   Fe\,{\sc i} line at 5250.2\,\AA{} by a comparison of a pixel in a particularly bright BP (red lines)
   with a pixel located in a much less bright BP (blue lines). Panel~(a) exhibits the vertical temperature
   stratification as a function of $\log(\tau)$ at 5000\,\AA{} and panel~(b) the magnetic field strength
   stratification. Panels~(c) and (d) depict the Stokes~$I/I_{\rm{QS}}$ and $V/I_{\rm{QS}}$ signals from
   the original MHD data at the twelve IMaX L12-2 scan positions in the line.}
   \label{FigTempEffect1}
   \end{figure*}

   Even at the original resolution of the simulations the average $\langle p_{\rm{circ}} \rangle$ is only
   6.66\,\% and only 2.6\,\% of all BPs reach the 10.71\,\% retrieved from the 1\,kG HSRASP atmosphere.
   This discrepancy can be explained by the high temperature sensitivity of the Fe\,{\sc i}
   line at 5250.2\,\AA{}, as illustrated in Fig.~\ref{FigTempEffect1}, where two pixels taken from the
   simulation data are compared. The pixel colored in blue belongs to a faint BP with a low brightness
   in the CN band, but one that is still identified as a BP, while the pixel colored in red is part of
   a BP with a high contrast. Panel~(a) of Fig.~\ref{FigTempEffect1} shows the vertical temperature
   stratification, where the atmospheric height is given in logarithmic units of $\tau$ (continuum
   optical depth at 5000\,\AA{}). In the middle photosphere (where the spectral line is mainly
   formed), the red pixel's temperature is about 800\,K higher than the temperature of the blue pixel,
   but in the lower and upper photosphere both temperatures are almost the same. The vertical stratification
   of the magnetic field strength is displayed in panel~(b). In the middle photosphere, e.g.,
   at $\log(\tau)=-2$, the blue pixel reveals a field strength of roughly 650\,G, while the red pixel
   has a field strength of 1200\,G. As expected for the flux-tube model of BPs, at $\log(\tau)=0$
   both pixels show a magnetic field stronger than 1\,kG that is nearly vertically oriented (the
   height averaged field inclinations at the two pixels are $8.9^{\circ}$ and $7.6^{\circ}$).
   The original Stokes~$I/I_{\rm{QS}}$ signals of the twelve scan positions of the 5250.2\,\AA{} line
   are shown in panel (c). The continuum intensity near 5250\,\AA{} of the red pixel is higher than
   $I_{\rm{QS}}$, while it is lower in the blue pixel. The most striking feature of the figure is the
   minute line depth of the 5250.2\,\AA{} line in the red pixel. This is partly due to the large Zeeman
   splitting caused by the large field strength in this pixel, but even more to the strong
   temperature sensitivity. The temperature sensitivity originates not just in the increased ionization
   of iron as the temperature is raised, but also from the excitation potential of the lower level of
   this line of only 0.12\,eV. The line weakening is also conspicuous in the Stokes~$V/I_{\rm{QS}}$
   profiles of panel~(d). Although the red pixel has a higher magnetic field strength, its circular
   polarization degree of 1.8\,\% is much smaller than that of the blue pixel, 6.8\,\%. The ratio
   1.8/6.8 is larger than the ratio of the line depths due to the contribution of the magnetic field
   which is stronger for the red pixel \citep[obviously Zeeman saturation,][is not complete, so that a
   residual Zeeman sensitivity of the Stokes~$V$ amplitude is present]{Stenflo1973}.

\begin{figure}
   \centering
   \includegraphics[width=\linewidth]{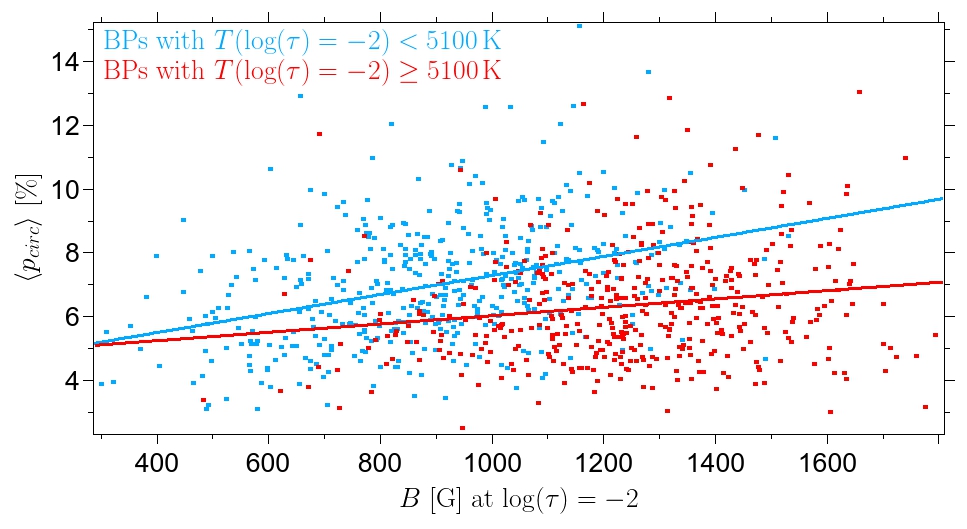}
   \caption{Circular polarization degree versus magnetic field strength at $\log(\tau)=-2$ as
   retrieved for the 898 BPs from the original 30\,G MHD data. Pixels with $T \ge 5100$\,K are colored
   in red, cooler pixels in blue. The solid lines are linear regressions.}
   \label{FigTempEffect2}
   \end{figure}

   The temperature effect is statistically relevant for the BPs in general. To show this, in
   Fig.~\ref{FigTempEffect2} we plot the circular polarization degree versus the magnetic field
   strength at $\log(\tau)=-2$ (peak values for both quantities) for all 898 BPs in the original
   30\,G MHD data. At $\log(\tau)=-2$ these BPs have a mean temperature of 5100\,K, and all BPs
   hotter than the mean temperature are colored in red, the cooler ones in blue. The solid lines
   are the linear regressions of these two BP classes. We recognize that for a given
   field strength, the BPs having a higher temperature show a weaker $\langle p_{\rm{circ}} \rangle$
   than cooler BPs, although the scatter is large, and there is some overlap. In addition, the hotter
   features tend to have stronger fields.

   Since the Stokes~$V$ normalization to $I_i$ chosen by \citet{Riethmueller2010} could enhance
   the effect of the line weakening of $\langle p_{\rm{circ}} \rangle$, we also considered other
   formulae to analyze the BP polarization or field strength, respectively: 

   \begin{equation}\label{Eq_Mean_CPc}
   \langle p_{\rm{circ}}^{\rm{QS}} \rangle = \frac{1}{12 I_{\rm{QS}}} \sum_{i=1}^{12}{\left| V_i \right|}~,
   \end{equation}

   \begin{equation}\label{Eq_Mean_Blos}
   \langle B_{\rm{LOS}} \rangle = \frac{1}{N} \sum_{i=1}^{N}{\frac{4 \pi c m_e}{e \lambda_0^2 g} \left| \frac{V_i}{(\rm{d}I/\rm{d}\lambda)_i} \right|}~.
   \end{equation}
   The scatterplot for $\langle p_{\rm{circ}}^{\rm{QS}} \rangle$ (not shown) exhibits the same
   qualitative behavior as that for $\langle p_{\rm{circ}} \rangle$ (shown in Fig.~\ref{FigTempEffect2}),
   so that the temperature effect cannot be reduced by this widely used normalization for the circular
   polarization calculation.

\begin{figure}
   \centering
   \includegraphics[width=\linewidth]{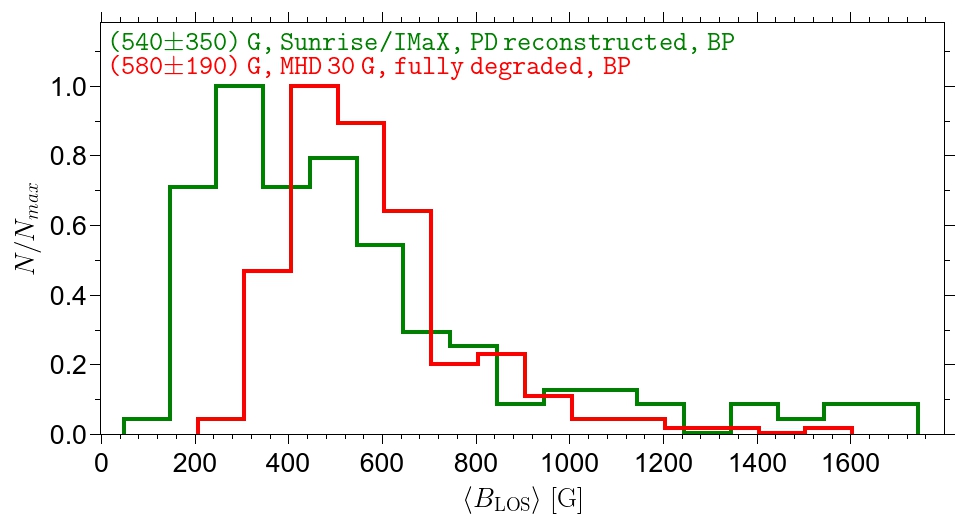}
   \caption{Histograms of the peak value of the BP field strength calculated via the weak field
   approximation (see main text). The green line shows the observed BPs and the red line displays the BPs
   from the degraded 30\,G MHD data.}
   \label{FigBPMean_BlosHist}
   \end{figure}

   Equation~(\ref{Eq_Mean_Blos}) is based on the weak field approximation which holds if the Zeeman
   splitting is much smaller than the line width \citep[e.g.,][]{LandiDeglInnocenti2004}. The derivative of
   Stokes~$I$ was determined from the Gaussian fit of the Stokes~$I$ profile. The values $c$, $m_e$,
   and $e$ have the usual meaning, $\lambda_0 = 5250.2$\,\AA{} is the reference wavelength, and $g = 3$
   is the Land\'e factor of the line. To avoid division by zero, the sum in Eq.~(\ref{Eq_Mean_Blos})
   was only calculated over the $N$ scan positions with $\left| \rm{d}I/\rm{d}\lambda \right| > 3 \sigma$,
   where $\sigma$ is the ratio of the Stokes~$I$ noise level and the scanning step size. A scatterplot
   of $\langle B_{\rm{LOS}} \rangle$ versus the magnetic field strength at $\log(\tau)=-2$ (not shown)
   revealed that Eq.~(\ref{Eq_Mean_Blos}) very significantly reduces the effect of the line weakening.
   Histograms of the $\langle B_{\rm{LOS}} \rangle$ values obtained from applying Eq.~(\ref{Eq_Mean_Blos}) to
   observational and degraded synthetic data are plotted in  Fig.~\ref{FigBPMean_BlosHist}. Similar
   to $\langle p_{\rm{circ}} \rangle$ (Fig.~\ref{FigBPMean_CPHist}), $\langle B_{\rm{LOS}} \rangle$
   shows a long tail of stronger fields which is more pronounced for the observational data. Even under
   this approximation, only 9.9\,\% of the observed BPs and 3.6\,\% of the synthetic BPs yielded
   $\langle B_{\rm{LOS}} \rangle$ values higher than 1\,kG.

\subsection{Properties of simulated bright points}
   In sections \ref{CompBPs} and \ref{ExplWeakBPs} we showed that the MURaM MHD simulations reproduce
   the properties of the observed BPs reasonably well. Therefore we can obtain a better understanding
   of the physical phenomena underlying BPs by analyzing BP properties in undegraded simulations.
   In the following we used the 898 BPs detected in the undegraded CN images.
   It is this set of BPs that underlies Figs.~\ref{FigBPBfieldHist}~to~\ref{FigScatter}.

\subsubsection{Magnetic field strength and inclination}
\begin{figure}
   \centering
   \includegraphics[width=\linewidth]{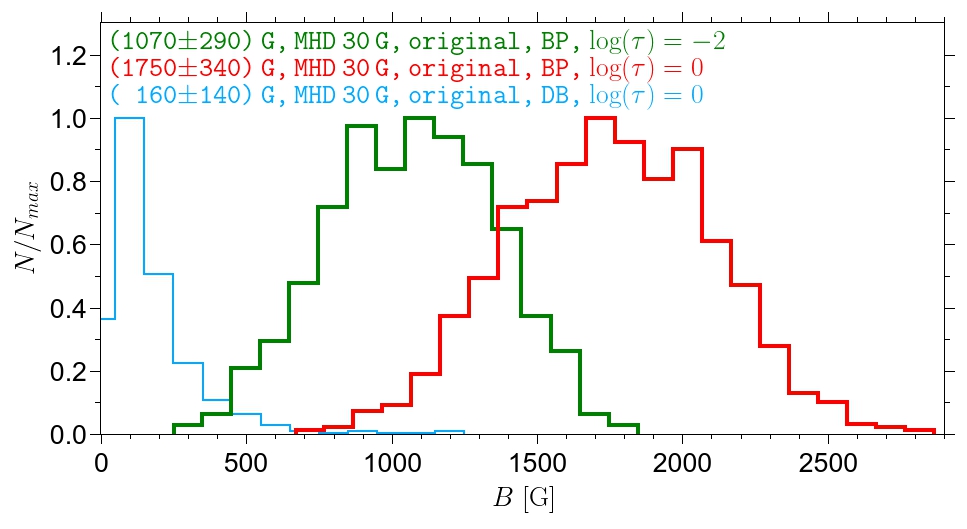}
   \caption{Histograms of the magnetic field strength of the BPs in the middle photosphere (green line),
   and of the BPs (red line) and the DB (blue line) in the lower photosphere as retrieved from the undegraded
   30\,G MHD data. Mean values and their standard deviations are given in the text labels.}
   \label{FigBPBfieldHist}
   \end{figure}

   Figure~\ref{FigBPBfieldHist} shows histograms of the BP peak magnetic field strength taken directly
   from the 30\,G MHD simulations. The peak values were determined as the maximum field strength at a
   given optical depth within a BP. Owing to lateral force balance, the decreasing gas pressure with height,
   and the need for magnetic flux conservation, the flux tubes expand, so that the BP field strength drops
   from an average of 1750\,G at $\log(\tau)=0$ to 1070\,G at $\log(\tau)=-2$. At optical depth unity,
   only 14 of the considered 898 BPs had a field strength lower than 1000\,G, i.e., 98\,\% of the BPs
   were in the kilogauss range, the strongest BP field was found to be 2825\,G, the weakest one
   was 721\,G. On average, the peak field strength in the DB is an order of magnitude lower than in the BPs.

\begin{figure}
   \centering
   \includegraphics[width=\linewidth]{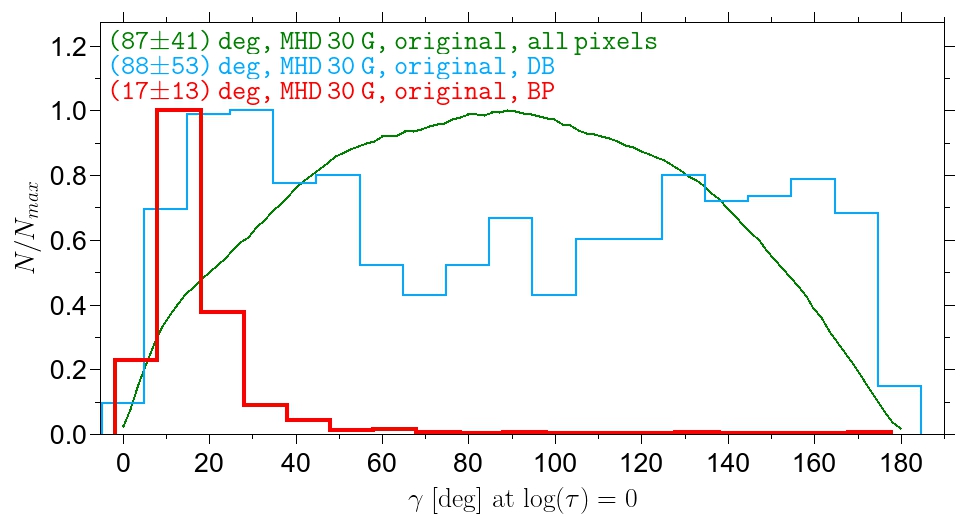}
   \caption{Histograms of the magnetic field inclination at optical depth unity of BPs, DB, and of all
   pixels. The color coding is the same as in the top panel of Fig.~\ref{FigBPOHHist}.}
   \label{FigBPGammaHist}
   \end{figure}

   Figure~\ref{FigBPGammaHist} displays the field inclinations (angle between the field vector and the
   surface normal) of the 30\,G MHD BPs at optical depth unity. The inclinations were averaged over all
   pixels covered by a BP at optical depth unity. The mean inclination\footnote{The undegraded data were
   noise-free and hence a magnetic field was found at every pixel, so that a field strength and inclination
   could also be given for each pixel.} at $\log(\tau)=0$ of the BPs is $17^{\circ}$, i.e., the BPs are almost vertical as
   expected from buoyancy considerations \citep{Schuessler1986}. Four of the 898 BPs showed inclinations
   greater than $90^{\circ}$, i.e., their magnetic field direction had been reversed from the initial
   condition of a homogeneous unipolar field. The field in the DB, in contrast, displays all possible
   inclination values, whereby almost vertical fields of either polarity are slightly preferred. The distribution
   of the field inclinations of all pixels is quite similar to the distribution of an isotropic field.

\subsubsection{Temperature}
\begin{figure}
   \centering
   \includegraphics[width=\linewidth]{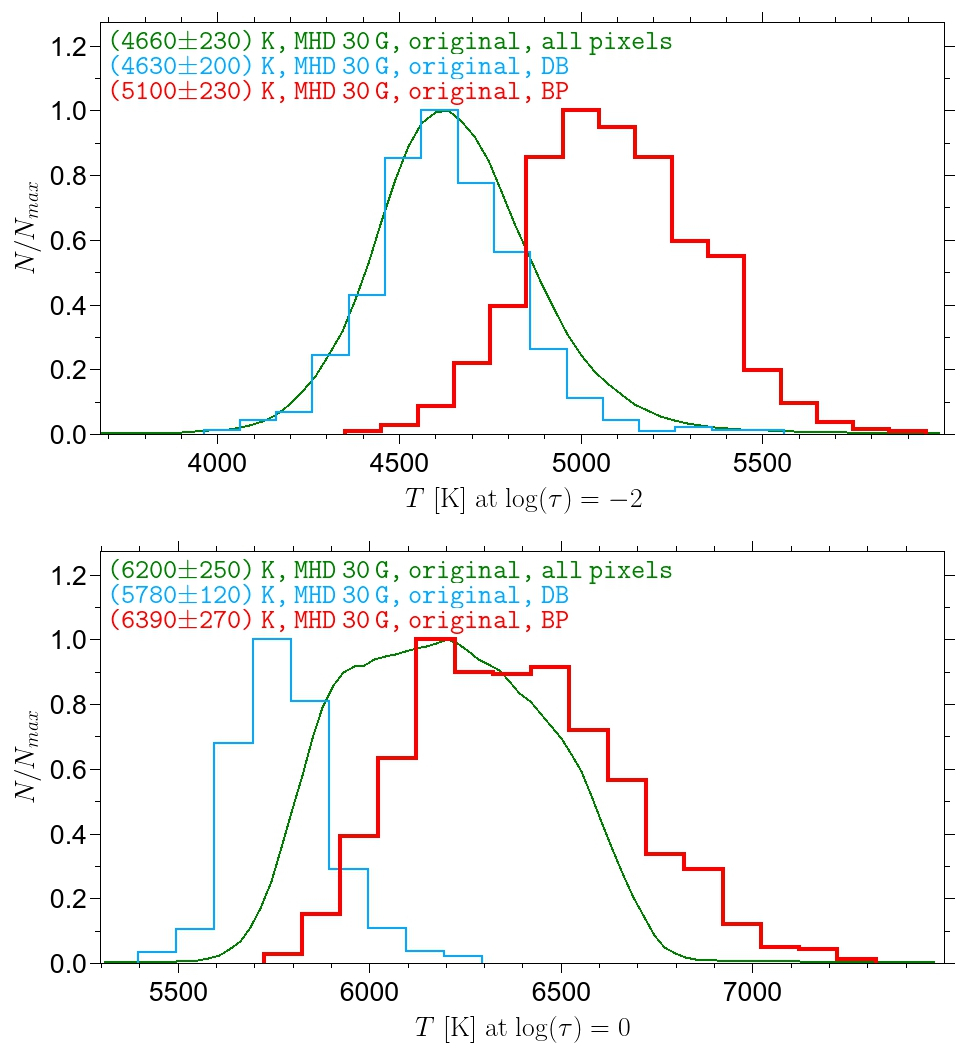}
   \caption{Histograms of the temperature for the simulated BPs (red lines), for the BPs' dark
   background (blue lines), and for all pixels in all frames (green lines). Mean values and their
   standard deviations are given in the text labels. The top panel shows the temperature at $\log(\tau)=-2$ and
   the bottom panel at optical depth unity (the $T$ scales are different).}
   \label{FigBPTempHist}
   \end{figure}

   In Fig.~\ref{FigBPTempHist} we compare the BP temperatures between the middle photosphere and the
   lower photosphere. The temperatures were averaged over all pixels of a BP as determined by the MLT
   algorithm (applied to the CN maps). The mean BP temperature in the middle photosphere was 440\,K higher
   and the mean DB temperature was 30\,K lower than the mean quiet-Sun temperature. In the lower
   photosphere, the mean BP temperature was 190\,K higher than the mean quiet-Sun temperature, while
   the dark background was 420\,K colder than the quiet Sun. The temperature gradient in the BP and
   in the DB were thus significantly lower than average.

\subsubsection{LOS velocity}
\begin{figure}
   \centering
   \includegraphics[width=\linewidth]{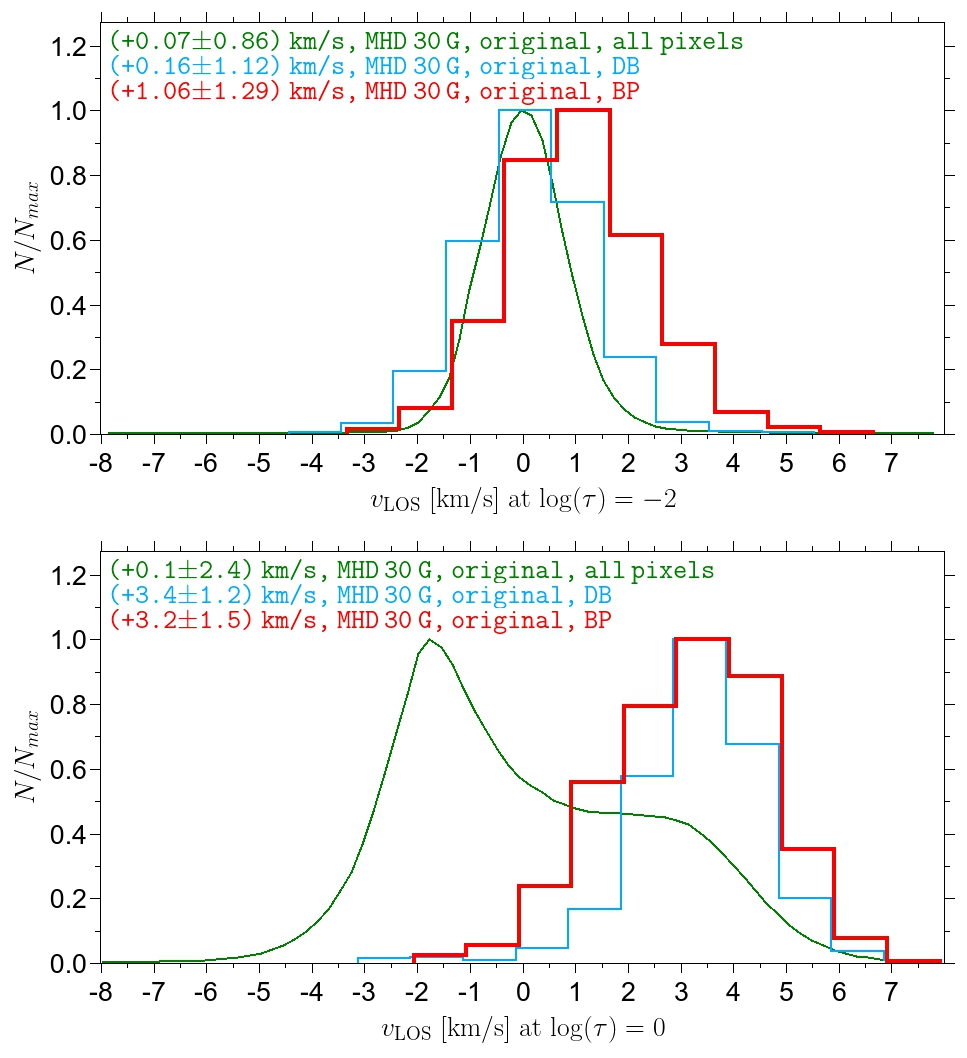}
   \caption{Same as Fig.~\ref{FigBPTempHist} for the LOS velocities (vertical component of the velocity
   vector taken directly from the MHD output).}
   \label{FigBPV_LOSHist}
   \end{figure}

   Histograms of the LOS velocity obtained directly from the MHD calculations are depicted in
   Fig.~\ref{FigBPV_LOSHist}. Again, the LOS velocities were averaged over all pixels of a BP as
   determined by MLT applied to 5250.4\,\AA{} continuum intensity images (for the same reason as in
   Sect.~\ref{BpObsVlos}). At an optical depth of $\log(\tau)=-2$ (top panel) the gas in the
   BP flows down at roughly 1\,km~s$^{-1}$, which is 0.9\,km~s$^{-1}$ faster than in the
   surroundings. At $\log(\tau)=0$ (bottom panel), the BPs showed a strong downflow of 3.2\,km~s$^{-1}$.
   This time the DB downflow was, on average, slightly stronger with 3.4\,km~s$^{-1}$. The histogram
   of all pixels (green line) clearly shows a superposition of two populations. The pixels from the
   interior of the granules formed the first population with an upflow of around $-2$\,km~s$^{-1}$ as
   the most numerous velocity. The second population corresponds to intergranular lanes with a
   3\,km~s$^{-1}$ downflow as the most common velocity.

\subsubsection{Correlations between BP properties}
   Finally, we were interested in correlations between BP properties found by looking at scatterplots and correlation coefficients.
   The spatial peak values of the field strength at optical depth unity, the CN intensity, as well as
   the 5250.4\,\AA{} continuum intensity were found to be only weakly correlated with the BP diameter
   (correlation coefficients are $0.17$, $0.19$, and $0.16$). The scatterplots (not shown) exhibited a
   slight increase in the three quantities for large BP diameters. A decrease in the intensities for
   very large diameters, expected for micropores and pores, could not be found in our study owing to
   the BP detection method, which looked for bright features and not for highly magnetized ones.
   This may also have to do with the rather limited range of BP diameters found in this very quiet region,
   with less than 4\% being broader than 200\,km (see the blue line in Fig.~\ref{FigBPDiameterHist}).
   
\begin{figure}
   \centering
   \includegraphics[width=\linewidth]{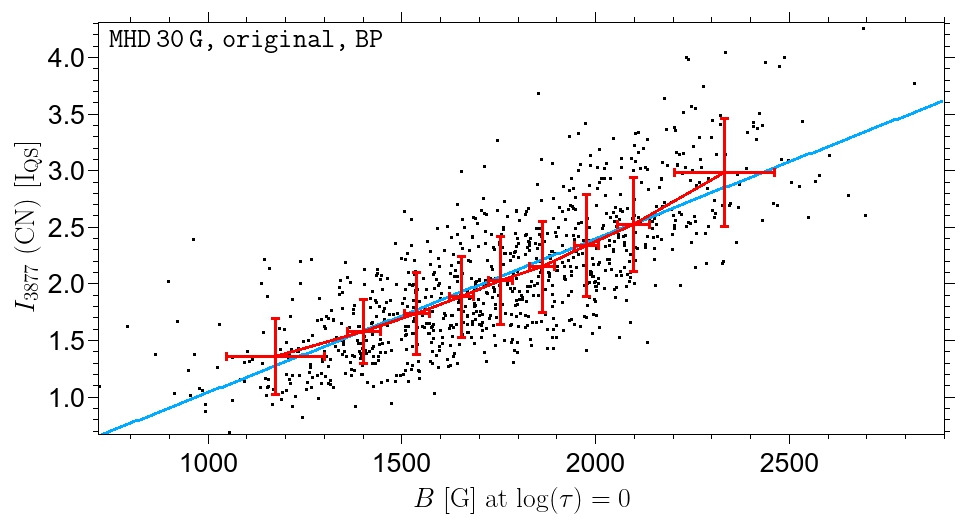}
   \caption{The black dots show a scatterplot of the BP peak CN intensity versus the magnetic field strength
   at optical depth unity. The red line connects binned values and the blue line is the linear regression.
   Error bars represent standard deviations.}
   \label{FigScatter}
   \end{figure}

   A clear relationship was found between the peak CN intensity and the peak magnetic field strength at
   optical depth unity with a correlation coefficient of $0.76$ (see Fig.~\ref{FigScatter}).
   The monotonicity over the whole range of field strengths is highlighted by binning points adjacent in $B$
   (red crosses in Fig.~\ref{FigScatter}). Again, a decrease in the intensity of very strong fields
   could not be found owing to the BP detection method which excluded micropores.

   Of some interest is the relation between the magnetic field inclination at optical depth unity and the peak CN intensity
   (not shown), which gave a moderate correlation coefficient of $-0.32$, implying that the more intense BPs are associated
   not just with a stronger field, but also with one that is on average more vertical. In this case all inclinations were transformed
   to the interval [0,90\,deg] to be independent of the field polarity and they were spatially averaged over the BP to get an estimate
   of the flux tube's orientation as a whole. From the darkest BPs ($0.68\,I_{\rm{QS}}$) to the brightest BPs ($4.3\,I_{\rm{QS}}$)
   the binning graph shows a nearly linear decrease in inclination from $23$\,deg to $5$\,deg.

%______________________________________________________________

\section{Discussion}
   The aim of this study is to learn more about magnetic BPs in the quiet solar atmosphere. To achieve
   this we combined high-resolution observations with realistic radiation MHD simulations. First we
   compared a row of parameters deduced from the observations (intensity, LOS velocity, line width,
   circular polarization degree) with their counterparts obtained from the simulations under conditions
   matching those of the observations as closely as possible (noise, spatial and spectral resolution,
   sampling, straylight, etc.). After establishing that the simulations give a reasonable
   description of the data, both in general and for the BPs in particular, we employed the
   simulations to deduce more about the BPs than can be gleaned from the observational data alone.

   We observed quiet-Sun regions at disk center with the balloon-borne observatory \sunrise{}.
   Photometric data at 3118\,\AA{} and 3877\,\AA{}, as well as spectropolarimetric data at twelve
   wavelengths in and around the Fe\,{\sc i} line at 5250.2\,\AA{} were acquired quasi simultaneously.
   Compared to high-resolution observations with other telescopes, we benefited from the following
   advantages of \sunrise{}: a) The Sun was observed not only in the visible but also in the near UV.
   b) The PSF was measured during the observations, so that the influence of the central obscuration
   by the secondary mirror, the spiders, and the low-order aberrations like defocus, coma, and
   astigmatism were known and we did not have to rely on a theoretical PSF. c) A stray light analysis
   was possible owing to observations of the solar limb. d) The data were practically free of seeing
   effects. Therefore, it was possible to carefully determine the degradations that were acting
   during the observations and to apply them to the MHD simulations. In contrast to other studies, which
   worked with opacity distribution functions \citep[ODFs, e.g.,][]{Danilovic2008,Afram2011}, we applied
   full spectral line syntheses, for the first time also to the OH band at 3118\,\AA{}, in order to
   compute observables to be compared with the \sunrise{} data. Our study concentrated on a comparison
   of the following BP properties at disk center: diameters, intensity in the visible and near UV,
   LOS velocity, polarization degree, and Stokes~$V$ asymmetry; other studies typically only focussed on
   center to limb variations of intensity histograms \citep{WedemeyerBoehm2009,Afram2011}.

   A reasonable match between the observations and the degraded simulations was found for the intensity
   histograms of all pixels for all three considered wavelength ranges, as well as for the LOS velocities
   and the circular polarizations for an initial field strength of 30\,G averaged over the simulated box,
   although some parameters indicated a better match with simulations containing more magnetic flux.
   We note that the multitude of considered observables strongly constrains the problem. The tuning of
   any degradation parameter to reach a better match with a particular observational quantity would
   significantly increase the mismatch with other observables.
   The intensity histograms of the undegraded 30\,G simulations showed a superposition of two populations
   which was greatly weakened by the degradation and which was not found in the observations. The two
   populations were also clearly present in comparable simulations with the Stagger and $\rm{CO^{5}BOLD}$
   code \citep{WedemeyerBoehm2009,Beeck2012}. \citet{Afram2010} also used the MURaM code and found the
   same superposition for their undegraded 0\,G and 50\,G simulations, but not for their 200\,G data.
   We can confirm this result by our intensity histograms of the degraded MHD data for varying mean
   flux densities.

   The degradation of our MHD data reduced the rms contrast at 3118\,\AA{} from 32.4\,\% to 21.1\,\%,
   at 3877\,\AA{} from 30.8\,\% to 20.5\,\% and at 5250.4\,\AA{} from 22.1\,\% to 12.1\,\%. With the help
   of ODFs, \citet{Hirzberger2010} calculated an rms contrast of 28.3\,\% at 3118\,\AA{} and of 23.9\,\%
   at 3877\,\AA{}. In both cases, complete spectral line syntheses led to significantly higher contrasts
   than with the ODF method, although the smaller grid size of our simulation may also
   have contributed. For the CN band at 3877\,\AA{}, \citet{Hirzberger2010} also applied a full spectral
   line synthesis, but they used an older version of SPINOR with a limited number of wavelength points.
   They converted from geometrical heights into optical depths with a different opacity package and
   they also used a different horizontal grid resolution of 20.8\,km. They report an rms contrast of
   25.3\,\% for the undegraded CN data while we found a value of 30.8\,\% with the improved SPINOR code.
   
   The deviation of the rms contrasts between degraded simulations and observations was 0.9\,\% at
   3118\,\AA{}, it was negligible at 5250.4\,\AA{}, at 3877\,\AA{} it was somewhat higher, but at 1.7\,\%
   the agreement was still remarkably good given the various sources of uncertainty, e.g., slightly varying
   image quality owing to the remaining pointing jitter of the gondola, inaccuracies of the spectral line
   synthesis \citep[incomplete spectral line lists, inaccurate atomic or molecular data, differences in the solar
   abundances found by, e.g.,][and neglected non-LTE effects]{Asplund2005}, and approximations
   involved in the radiative transfer calculations of the MHD simulation (only four opacity bins, LTE/no scattering,
   limited angular resolution). Given that the computed CN lines are clearly too strong even for the
   average quiet Sun (see Fig.~\ref{FigSynthSpectra}), we expect the inaccuracies of the spectral line
   synthesis to play a significant role in producing the discrepancies in the UV. A comparison between several MHD codes
   by \citet{Beeck2012} led to a variation of up to 1\,\% even in the bolometric intensity contrast.
   \citet{Hirzberger2010} found rms contrasts of 18.3\,\% and 20.1\,\% for two different \sunrise{}
   observations at 3877\,\AA{}, while our observation showed 18.8\,\%. Depending on the ratio of the
   inverse jitter frequency to the exposure time, different image qualities are possible for observations
   at different wavelengths. The exposure time of the 3877\,\AA{} images was by far the shortest one
   compared to the other wavelengths and thus this wavelength range was expected to have sharpest images.
   
   A comparison of our observed rms intensity contrasts with ground-based observations cannot be done easily
   owing to the large influence of seeing effects and atmospheric stray light on the image quality.
   Hence, we compare with data of the space-borne Solar Optical Telescope onboard the Hinode satellite.
   \citet{Mathew2009} determined the PSF of the Broadband Filter Imager with the help of Mercury transit images
   at different wavelengths and deconvolved quiet-Sun images. For the violet CN band they reported an rms contrast
   of 21.8\,\% and of 15.8\,\% for the green continuum at 5550\,\AA{} (which is a wavelength relatively close
   to our 5250.4\,\AA{} Fe\,{\sc i} continuum). These values are somewhat higher than ours, 18.8\,\% and
   12.1\,\%, probably because their PSF also removed the stray light while it was still present in our data.

   We found that the various degradation steps considerably influenced the shape of the histograms
   of the parameters. In particular, the stray light influenced histograms of all quantities and often
   it provided the most important contribution to the degradation. This was also found by
   \citet{WedemeyerBoehm2009}. Additionally, for the line width histogram a good knowledge of the
   spectral PSF of the instrument is needed. The circular polarization histograms depended strongly
   on the noise level of the Stokes~$V$ images, while the noise level of the Stokes~$I$ images
   as well as the other degradation steps played only a minor role (due to the high signal-to-noise
   ratio in Stokes~$I$).

   Stray light may also contribute to the last remaining discrepancy between the data and the simulations,
   the larger scatter of the observed 5250.2\,\AA{} line widths than in the simulations. We used the
   continuum images of the IMaX limb observations, but the IMaX stray light may depend on the
   wavelength within the spectral line. Because of the sensitivity of the histogram shapes to stray light,
   we believe that such a wavelength dependence could be a good candidate to explain the larger
   scatter in the observed line widths. Another possibility is the influence of the evolution of the
   granulation during the relatively long acquisition time of IMaX of almost 32\,s, which may also
   change the line width histogram. Another possibility, insufficient turbulence in the MHD simulations,
   would only explain the missing tail of high line widths but not the missing low widths.

   The good match between observation and simulation (only a moderate deviation remained for the scatter
   of line widths) enhanced our trust in the simulations, so that we used them to probe BP properties.
   We found a BP number density of 0.05 BPs per $\rm{Mm}^2$ in our observations which is at the lower end
   of the wide spread of values found in the literature: \citet{Muller1984} reported 0.04 BPs per
   $\rm{Mm}^2$, 0.12 BPs per $\rm{Mm}^2$ were found by \citet{Bovelet2008}, 0.3 BPs per $\rm{Mm}^2$ by
   \citet{SanchezAlmeida2004}, and even 0.85 BPs per $\rm{Mm}^2$ by \citet{SanchezAlmeida2010}. All
   these were quiet-Sun studies. Active-region data analyzed by \citet{Berger1995} gave 0.37 BPs per
   $\rm{Mm}^2$. Our BP number density is larger than the 0.03 BPs per $\rm{Mm}^2$
   obtained by \citet{Jafarzadeh2013} from Ca\,{\sc ii}~H observations made by \sunrise{}, but this is
   likely because they restricted their study to features narrower than 0\carcsec{}3.
   
   The effective BP diameter was, on average, 334\,km in our observations using MLT and 220\,km from the
   FWHM of the average BP intensity profile, despite the fact that the MLT boundaries of the BPs are defined
   by a threshold of 50\% of the local min-max intensity range. We note that for an idealized brightness
   structure of a rotationally symmetric Gaussian shape MLT provides a diameter that is equal to the
   FWHM value of the mean intensity profile, but already a Gaussian with an elliptical cross-section
   leads to different results. This demonstrates the sensitivity of the diameter determination on the
   technique. Most other studies reported BP diameters consistent with our value obtained from
   the average BP profile; for example, \citet{Berger1995} found a mean FWHM intensity diameter of 250\,km by
   taking the smallest dimension across the BP features (whereas we obtained an effective diameter).
   \citet{SanchezAlmeida2004} fitted the minor and major axes of the BPs with average values of
   135\,km and 230\,km, respectively. \citet{Utz2009} reported a decrease in the mean BP diameter
   from 218\,km to 166\,km by doubling the spatial sampling of data from the Hinode telescope.
   \citet{Crockett2010} found a distribution that peaked at an effective BP diameter of 240\,km.
   Combining our observed BP number density with the mean BP diameter results in a 0.5\% fraction of
   the solar surface that is covered by quiet-Sun BPs. This value is only half of the 1\% found by
   \citet{Bonet2012} using reconstructed disk center data from the ground-based 1~m Swedish Solar Telescope.
   
   The observed BP sizes found in the literature are generally larger than the 129\,km obtained by MLT directly
   from the undegraded simulations. We note that this value was reduced by 20\% when going from a grid size
   of 20\,km to 10\,km, suggesting that the claim of \citet{Crockett2010} that they had resolved
   all BPs, based on a comparison of observations with MURaM MHD simulations with a grid size of
   25\,km, was premature. The fact that nearly half of our undegraded synthetic BPs have a diameter
   smaller than 130\,km (see the blue line in Fig.~\ref{FigBPDiameterHist}) casts doubts on the
   claim of \citet{Wiehr2004} that there is a true deficit of such small BPs in the observations
   which does not stem from instrumental artifacts.
   
   A direct comparison of the BP number density and BP diameter from studies made at different
   telescopes is not straightforward because both quantities depend strongly on the image
   degradation (i.e., we cannot say much about the underlying true size distribution),
   on the mean vertical flux density of the observed region, on the method employed to identify
   the BPs, and possibly on the phase of the solar cycle. For our simulated BPs we found that
   degrading the data to the \sunrise{} resolution decreased the BP number density by a factor
   of 3.2, while the BP diameter was increased by a factor of 2.6. An increase of the mean
   vertical flux density from 30\,G to 200\,G led to a 3.1 times higher BP number density.
   We note that our manual detection method tends to consider larger bright patches (associated
   with larger concentrations of magnetic flux) as single features whereas other techniques
   may identify them as chains or clusters of multiple BPs, which might also be a reason
   that we found a lower BP number density and a higher BP diameter, at least when applying MLT
   to the individual BPs. Another reason for the relatively low number of BPs could be that \sunrise{}
   observed a very quiet region in the deepest part of the last activity minimum. \citet{Foukal1991}
   and \citet{Meunier2003} found evidence that the number of quiet-Sun BPs is correlated with
   the solar cycle. However, \citet{Muller1984} obtained an anticorrelation from their observations.

   Asymmetries of the Stokes~$V$ profiles provide insights into the structuring of magnetic fields
   along the LOS \citep{Solanki1988,GrossmannDoerth1988,SanchezAlmeida1989,Steiner2000,LopezAriste2002,Shelyag2007,MartinezGonzalez2012}.
   The rapidly increasing asymmetries with distance from the mean BP core are indicators of the flux-tube
   canopy. Flux tubes expand with height owing to pressure balance and magnetic flux conservation.
   \citet{MartinezGonzalez2012} also analyzed IMaX L12 data of a single network element and found
   evidence for a canopy. By studying a statistical ensemble of BPs, mostly consisting of internetwork
   elements, we were able to show that the spatial distribution of the asymmetry is consistent with the presence
   of a canopy, thus further supporting the model of \citet{GrossmannDoerth1988} and \citet{Solanki1989}.

   For a deeper understanding of the BP phenomenon, we analyzed properties of the simulated BPs in greater
   detail. An important finding is that basically all magnetic BPs identified on the basis of their
   brightness properties in the simulations are associated with kilogauss fields, with the average $B$ at
   $\log(\tau)=0$ being 1760\,G, which drops to 1070\,G at $\log(\tau)=-2$. This is in contrast to the
   observations, which gave 540\,G (from the weak field approximation). This is comparable with the values obtained
   from the simulations after all the effects of the instrument have been taken into account. This difference,
   as well as that between the BP sizes before and after applying instrumental effects, indicates that
   many BPs are not fully resolved in our data set. We note that because of the higher resolution of \sunrise{}
   data taken at other times, more features are expected to be resolved there \citep[e.g.,][]{Lagg2010}.
   We also note that instrument degradations imply that we are missing nearly two thirds of all BPs in the
   quiet Sun in our data. The fact that the BP contrast increases quickly with $B$, but does not depend
   strongly on diameter, suggests that for these slender features the field strength is the main driver
   of flux-tube brightness.

   The histogram of the inclination of the magnetic field vector exhibited a nearly vertical magnetic
   field for most of the BPs, which is to be expected for kilogauss fields and not too strong horizontal
   flows \citep{Schuessler1986}. This result is in good agreement with the observational findings of
   \citet{Jafarzadeh2014a}, obtained by comparing the positions of BPs recorded at different heights by
   \sunrise{}. Hence this is another point in which \sunrise{} data and MHD simulations agree.

   A significant difference between the middle and lower photosphere was found for the vertical
   velocities of the simulated BPs. While the BPs showed on average a downflow of 3.2\,km~s$^{-1}$
   in the lower photosphere, the downflow was reduced to 1.06\,km~s$^{-1}$ in the middle photosphere.
   A decrease of velocity with height was also found by \citet{BellotRubio1997} from the inversion of
   asymmetric Stokes~$V$ profiles observed in plage regions. The deeper we look into the atmosphere
   the stronger are the downflows of the BPs, so that it would be interesting to observe BPs
   spectropolarimetrically in a spectral line that is formed deep in the photosphere, e.g., the
   C\,{\sc i} line at 5380\,\AA{}, which we expect to show BP downflows that are stronger than the
   ones found with \sunrise{}. These strong flows in the deep layers of photospheric magnetic elements
   raise the question of the origin of the mass. Either it diffuses into the magnetic features across
   field lines, which runs counter to the estimates of \citet{Hasan1985}, or the lifetimes of BPs,
   i.e., kilogauss features are rather short, or the plasma with the strong field is continually mixing with
   relatively field-free plasma in the immediate surroundings of the magnetic elements. This last process
   may be related to the vortices found in the simulations around magnetic elements by, e.g., \citet{Moll2011},
   and observationally by \citet{Bonet2010} and \citet{WedemeyerBoehm2012}.

\section{Conclusions}
   We have compared high-resolution \sunrise{} data in three spectral bands with three-dimensional
   radiative MHD simulations and found that the two agree remarkably well in most areas, as long as
   all instrumental effects that degrade the data are properly introduced into the simulations as
   well. This represents a stringent test of the simulations, since we consider many more parameters
   than just intensities. In addition, we consider both the entire FOV as well as BPs separately.

   We showed that although most of the BPs are weakly polarized in the observational data
   \citep[see also][]{Riethmueller2010} they correspond to magnetic elements with kilogauss fields.
   The small signals can be partly explained by a combination of thermal weakening of the temperature
   sensitive Fe\,{\sc i} 5250.2\,\AA{} line, spatial smearing due to residual pointing jitter, and
   instrumental stray light. In the original simulations 98\,\% of the BPs are almost vertically
   oriented magnetic fields in the kilogauss range, which, together with our findings about the
   asymmetries in the circular polarization signals, confirms the physical model of magnetic flux
   concentrations as evacuated and laterally heated structures that expand with height
   \citep{Spruit1976,Deinzer1984}. The field strength in the magnetic elements is found to be the
   quantity with the largest effect on its brightness.

   Magneto-hydrodynamical simulations with a horizontal cell size of 20\,km or larger are widely used in the literature
   \citep{Afram2010,Afram2011,Hirzberger2010,OrozcoSuarez2008,Roehrbein2011,Tritschler2006,WedemeyerBoehm2009}.
   We found a reduction of the effective BP diameter of 20\% and a near doubling of the BP number
   density by doubling the horizontal grid resolution. We cannot rule out that MHD simulations with
   a horizontal cell size lower than the 10\,km used here will give yet other BP properties.

   The observations, in particular when taken together with the simulations, also indicate that
   phenomena are present in the BPs that are not yet understood. One of these is the average downflow velocity
   of 0.6\,km~s$^{-1}$ in the observed Stokes~$I$ profiles and 0.27\,km~s$^{-1}$ in the degraded
   synthesized line profiles. An even higher mean downflow velocity of 1.25\,km~s$^{-1}$ is obtained from the
   original synthesized line profiles, i.e., if we do not consider instrumental degradation effects.
   Such high universal downflows would lead to an evacuation of the gas by an
   order of magnitude within the time it takes the gas to flow down two scale heights,
   i.e., roughly 200\,km. At 1.25\,km~s$^{-1}$ this will take place within 160\,s. Comparing with the
   mean lifetime of Ca\,{\sc ii}~H BPs of 673\,s in similar \sunrise{} data sets \citep{Jafarzadeh2013},
   this implies that the magnetic elements would be almost completely evacuated within a fraction of
   their lifetimes unless the gas is continuously replenished. An even stronger need for replenishment
   is present in the simulations, which show downflows of 3.2\,km~s$^{-1}$ at $\log(\tau)=0$ in
   BPs\footnote{The mean downflow velocity of 3.2\,km~s$^{-1}$ is obtained from the z components of the
   native MHD velocity vectors at optical depth unity, while the 1.25\,km~s$^{-1}$ mentioned above is
   calculated from Gaussian fits to the undegraded synthesized Stokes~$I$ profiles.}. Such
   flows would evacuate the magnetic element within 63\,s.

   The plasma can be replenished by gas flowing up along the opposite footpoints of loops that
   end in the BPs. As shown by \citet{Wiegelmann2010} at the \sunrise{} resolution in the quiet Sun,
   strong-field regions, such as BPs, are mainly connected to weak-field regions. Furthermore, most of these
   loops are rather low-lying, i.e., not reaching above the chromosphere. Along such loops a siphon
   flow from the footpoint with weaker field to that with the stronger field can take place
   \citep{Meyer1968,Montesinos1989}. Such a flow would produce a downflow in the BPs, as has been
   observed along the neutral line of an active region \citep{Rueedi1992,Degenhardt1993}.
   However, the simulations also show these downflows and they have a closed upper boundary through
   which the magnetic field of the BPs passes, but no flow is allowed to go through. Hence the flow must be
   replenished locally. We note that the observed average downflows of 0.6\,km~s$^{-1}$ are larger
   than the ones obtained from the degraded synthesized line profiles, 0.27\,km~s$^{-1}$, and we also note that the
   observed intensities are higher (see Figs.~\ref{FigBPOHHist}~and~\ref{FigBPCCTHist}). The difference
   may be due to missing siphon flows in the simulations, or just to the lower Reynolds number of the simulations
   compared with the real Sun. The lower the Reynolds number, the more intense the artificial viscous braking,
   which leads to lower velocities.

   The \sunrise{} images contain a few large and highly polarized BPs which were possibly part of
   network elements. Such network elements were not present in our simulations, presumably because
   the used simulation box is too shallow, which leads to a lack of larger scale flows accumulating magnetic flux.
   Usually, network elements are observed at the boundaries of supergranules and are possibly formed
   deeper in the convective zone \citep{Schuessler2013}. In a future study we will analyze MHD simulations
   with a computational box large and deep enough to contain one or more supergranules.

   \begin{acknowledgements}
   We thank Robert Cameron for the efficient MURaM lessons.
   The German contribution to \sunrise{} is funded by the Bundesministerium
   f\"{u}r Wirtschaft und Technologie through Deutsches Zentrum f\"{u}r Luft-
   und Raumfahrt e.V. (DLR), Grant No. 50~OU~0401, and by the Innovationsfond of
   the President of the Max Planck Society (MPG). The Spanish contribution has
   been funded by the Spanish MICINN under projects ESP2006-13030-C06 and
   AYA2009-14105-C06 (including European FEDER funds). The HAO contribution was
   partly funded through NASA grant number NNX08AH38G. This work was partially
   supported by the BK21 plus program through the National Research Foundation (NRF)
   funded by the Ministry of Education of Korea.
   \end{acknowledgements}

%______________________________________________________________

\end{document}